\documentclass[11pt]{article}

\usepackage[margin = 2.50cm]{geometry}

\usepackage{microtype}%if unwanted, comment out or use option "draft"
\usepackage{epsfig}
\usepackage{microtype}
\usepackage{graphicx}
\usepackage{enumitem}
\usepackage{epsfig,enumerate,amsmath,amsfonts,amssymb,amsthm,mathrsfs,ifpdf}
\usepackage{indentfirst,relsize}
\usepackage[numbers]{natbib}
\usepackage{setspace}
\usepackage{amsmath}
\usepackage[linesnumbered,ruled]{algorithm2e}
\usepackage{latexsym}
\usepackage{stackrel}
\usepackage[all]{xy}
\usepackage[usenames,dvipsnames]{pstricks}
\usepackage{pst-grad} % For gradients
\usepackage{pst-plot} % For axes
\usepackage{xspace}

\newcommand{\size}[1]{\left| #1 \right|}

\newcommand{\remove}[1]{}

\newcommand{\R}{\mathbb{R}}

\newcommand{\cI}{\mathcal{I}}

\newcommand{\cS}{\mathcal{S}}

\newcommand{\cV}{\mathcal{V}}
\newcommand{\cC}{\mathcal{C}}

\newcommand{\cA}{\mathcal{A}}
\newcommand{\cB}{\mathcal{B}}

\newcommand{\cP}{\mathcal{P}}
\newcommand{\cK}{\mathcal{K}}
\newcommand{\cR}{\mathcal{R}}

\newcommand{\cO}{\mathcal{O}}
\newcommand{\cCH}{\mathcal{CH}}
\newcommand{\cH}{\mathcal{H}}

\newcommand{\eps}{\epsilon}
\newcommand{\p}{\mbox{Pr}}
\newcommand{\geo}{D_g}
\newcommand{\stgeo}{D_g^*}
\newcommand{\col}{D_c}
\newcommand{\stcol}{D_c^*}
\newcommand{\alphacol}{D_{c,\alpha}}
\newcommand{\halfcol}{D_{c,\frac{1}{2}}}

\newcommand{\ceil}[1]{{\lceil{#1}\rceil}}
\newcommand{\floor}[1]{{\lfloor{#1}\rfloor}}

\newcommand{\comments}[1]{\textcolor{blue}{\bf{#1}}}
\newcommand{\convhull}{\textsc{convex-body}\xspace}
\newcommand{\geoqpoly}{\textsc{geom-query-conv-poly}\xspace}
\newcommand{\geoqchain}{\textsc{geom-query-conv-chain}\xspace}
\newcommand{\coordmax}{\textsc{skyline}\xspace}
\newcommand{\simpc}{\textsc{simple-curve}\xspace}
\newcommand{\klee}{\textsc{Klee's measure}\xspace}
\newcommand{\closep}{\textsc{closest-pair}\xspace}
\newcommand{\aclosep}{\textsc{approx-closest-pair}\xspace}
\newcommand{\rankq}{\textsc{rank-query}\xspace}
\newcommand{\nnq}{\textsc{nearest-neighbor-query}\xspace}
\newcommand{\gdisc}{\textsc{geometric-discrepancy}\xspace}
\newcommand{\disc}{\textsc{star-geometric-discrepancy}\xspace}
\newcommand{\cdisc}{\textsc{color-discrepancy}\xspace}
\newcommand{\stcdisc}{\textsc{star-color-discrepancy}\xspace}
\newcommand{\disj}{\textsc{Disj}\xspace}
\newcommand{\lp}{\textsc{LP}\xspace}
\newcommand{\ind}{\textsc{Index}\xspace}
\newcommand{\eql}{\textsc{Equality}\xspace}

\theoremstyle{plain}
\newtheorem{theo}{Theorem}%[section]
\newtheorem{lem}[theo]{Lemma}

\newtheorem{coro}[theo]{Corollary}

\newtheorem{cl}[theo]{Claim}
\theoremstyle{definition}

\newtheorem{rem}{Remark}
\newtheorem{obs}[theo]{Observation}
\newtheorem{fact}[theo]{Fact}

\title{On the streaming complexity of fundamental geometric problems}
\author{
Arijit Bishnu
\footnote{
Indian Statistical Institute, Kolkata, India
}
\and
Arijit Ghosh
\footnotemark[1]
\and
Gopinath Mishra
\footnotemark[1]
\and
Sandeep Sen
\footnote{
Indian Institute of Technology Delhi, India  
}
}
\begin{document}

\maketitle
\begin{abstract}
In this paper, we focus on lower bounds and algorithms for some 
basic geometric problems in the one-pass (insertion only) streaming model. The problems considered
 are grouped into three categories --- 
\begin{itemize}
\item[(i)]
Klee's measure

\item[(ii)]
  Convex body approximation, geometric
query, and 

\item[(iii)] 
Discrepancy
\end{itemize}

Klee's measure is the problem of finding the area of the union of
hyperrectangles. Under convex body approximation, 
we consider the problems 
of convex hull, convex body approximation, linear programming (\lp) in
fixed dimensions. The results for convex 
body approximation implies a property testing type result to find if a
query point lies inside a convex polyhedron. Under discrepancy, we consider
both the geometric and combinatorial discrepancy. For all the problems
considered, we present (randomized) 
lower bounds on space. Most of our lower bounds are in terms of
approximating the solution with respect to 
an error parameter $\eps$. \remove{using distributional complexity or reductions from communication complexity} 
We provide approximation algorithms that closely match the lower bound on space for most of the problems.  
\end{abstract}
%\vspace{-0.2cm}

\section{Introduction} 
\label{sec:intro}

\noindent A {\em data stream} $\cP = \{p_1,\ldots, p_n\}$ is a sequence of data that can be read in increasing order of its indices $i$ ($i=1,
\ldots, n$) in one or more passes. In this paper, we 
consider the \emph{one-pass, insertion only streaming model}. 
For us, $\cP$ will be typically a set of points in $\R^d$. Only a sketch $\cS$, that is 
either a subset of $\cP$ or some information derived from it, can be stored; $\size{\cS} \ll \size{\cP}$. 
As a machine model, streaming has just the bare essentials. Thus, impossibility results, in
terms of lower bounds on the sketch size, becomes important.\remove{Historically, the streaming algorithms of Morris~\cite{Morris78}, Munro and
Paterson~\cite{MP78} and Flajolet et al.~\cite{Flajolet85} were precursors to the seminal work of Alon et al.~\cite{AlonM99} where the idea of lower bounds on space for approximating frequency moments was considered.} The seminal work of Alon et al.~\cite{AlonM99} introduced the idea of lower bounds on space for approximating frequency moments.
\remove{Subsequently, lower bounds on space usage of streaming algorithms have been a subject of intense research~\cite{IndykW10,CKS03,HenzingerRR98,IndykW03,Andrew14,Muthu05,Tim16,Woodruff04}.
}
The focus
on massive data applications has generated a lot of interest in streaming
algorithms and related lower bounds~\cite{GuhaM12,Muthu05,Tim16,Woodruff04}. 
In this paper, we try to 
push the frontiers of streaming in computational geometry by addressing 
fundamental problems like Klee's measure, convex body approximation, discrepancy, etc both in terms of lower bounds and matching algorithms . We 
also consider \emph{promise} problems and \emph{property testing} 
kind of results for some problems. \remove{In \emph{promise} problems, the input is promised to satisfy some properties.}

\subsection{Our computational model and notations}
\label{ssec:prob}

\noindent We will deal with points in $\R^d$ that 
can be represented as rationals with bounded bit precision. 
Thus, any point in our stream
$\cP$ comes from a universe of size $[N]^d$, where $[x]$ denotes
$\{1,\ldots,x\}$. Computations take place 
in a word RAM whose word size can hold a point, any input
parameter and a $\ceil{\log n}$-bit counter. The precision of the intermediate 
data generated is within a constant factor of the word 
size. The size of the stream $\size{\cP}=n$ is not 
known beforehand but standard techniques allow us to assume that wlog.\remove{The distance between two points $p$ and $q$ is denoted as $\delta(p,q)$
and the closest pair distance among the points of $\cP$ is denoted as
$\delta^*(\cP)$. $\delta^*(q,\cP)$ denotes $\min\limits_{p
\in \cP}\delta(q,p)$. $\cS$ will denote the sketch of $\cP$.} 

Let $\left[ p,q  \right]$ denote an interval between $p$ and $q$ in $\R$ and $\size{\left[ p,q  \right]}$, its length.
For a problem $P$, let ${\cal O}$ and ${\cal O}'$ be the optimal and an algorithm generated solution, respectively. By $\eps$-additive and $\eps$-multiplicative solutions to $P$, we mean $\size{{\cal O}'- {\cal O}} \leq \eps$ and $\size{{\cal O}'- {\cal O}} \leq \eps{\cal O}$,  respectively. 
A convex body $\cal K$ is said to be $\eps$-approximated by a convex body ${\cal K}'$ if $d_H({\cal K}, {\cal K}') \leq \eps$, where $d_H(\cdot, \cdot)$ denotes the Hausdroff distance. In our context,
 the diameter of $\cK$ is bounded by $1$. 
By $(\eps,\rho)$-additive solution, we mean $\eps$-additive solution that succeeds with probability at least $1-\rho$. Similarly, we define $(\eps,\rho)$-multiplicative solution. Also, by $(\eps,\rho)$-approximate solution, we mean $\eps$-approximate solution that succeeds with probability $\rho$. Typically $\rho = \frac{1}{3}$, unless stated otherwise.

In almost all the problems considered in this paper, the optimal solution lies in $\left[0,1\right]$ and we provide lower bound results for $\eps$-additive solution. Note that in such cases, the lower bound results for $\eps$-additive solutions are relatively stronger than their $\eps$-multiplicative counterparts. We also provide one-pass algorithms for $\eps$-additive solutions which can be converted into multi-pass algorithms for $\eps$-multiplicative solutions. This can be achieved easily because of the following theorem. Note that upper bound results for $\eps$-multiplicative solutions are relatively stronger than their $\eps$-additive counterparts.

\begin{theo}
\label{theo:add_mul}
Let $P$ be a problem whose optimal solution is ${\cal O}\in \left(0,1\right]$. Let $\cA$ be a one-pass alorithm that gives $\eps$-additive $(\eps \in (0,1))$ solution to $P$ using space $S(\eps)$. We can design an 
$O \left( \log \left( \frac{1}{\cO}\right) \right)$-pass algorithm $\cA'$ that gives $\eps$-multiplicative solution to $P$ and uses space $S\left(\eps \cdot \cO \right)$. 
\end{theo}
\begin{proof}
We design a $p$-pass algorithm $\cA'$ by invoking $\cA$ repeatedly. We fix $p$ later. In the $k$-th pass, $k \in [p]$, we do the following. 

We find an $\eps_k$-additive solution to $P$ using $\cA$, where $\eps_k=\frac{\eps}{2^{k}}$. Let $\cO_k$ be the corresponding output. Note that $\size{\cO_k-\cO}\leq \eps_k$. Assuming $\cO_k \geq \frac{1}{2^k}+\eps_k$, we can deduce the following.
\begin{eqnarray*}
\cO_k &\geq& \frac{1}{2^k}+\eps_k \\
\cO &\geq& \frac{1}{2^k} ~~(\because  \size{\cO_k-\cO}\leq \eps_k)\\
\eps \cdot \cO &\geq& \eps_k ~~(\because  \eps_k =\frac{\eps}{2^k})\\
\size{\cO_k-\cO}  &\leq& \eps \cdot \cO ~~(\because  \size{\cO_k-\cO}\leq \eps_k)
\end{eqnarray*}
  If $\cO_k \geq \frac{1}{2^k}+\eps_k$ holds in the $k$-th pass, then we return $\cO_k$ as the $\eps$-multiplicative solution to $\cP$.
  
  Recall that $\cA'$ is a $p$-pass algorithm. So, $\cO_p \geq \frac{1}{2^p}+\eps_p$ and  ${\cal O}_k < \frac{1}{2^k}+\eps_k$ 
  for each $1 \leq k < p$. Now we show that for $p = \ceil{\log \left(\frac{1 +2\eps}{{\cal O}}\right)}$, ${\cal O}_p \geq \frac{1}{2^p}+\eps_p$ holds. 
  \begin{eqnarray*}
  p &\geq& \log \left(\frac{1 +2\eps}{\cO}\right) \\
  \cO &\geq& \frac{1+2\eps}{2^p}\\
  \cO &\geq& \frac{1}{2^p}+2 \eps_p \\
   \cO_p &\geq& \frac{1}{2^p}+ \eps_p \\
  \end{eqnarray*}
   \remove{${\cal O}_p \geq \frac{1}{2^p}+\eps_p$when ${\cal O}-\eps_p \geq \frac{1}{2^p}+\eps_p$ as ${\cal O}_p \geq {\cal O} -\eps_p  $. ${\cal O}-\eps_p \geq \frac{1}{2^p}+\eps_p$ implies ${\cal O} \geq \frac{1+2\eps}{2^p}$. Hence, we can set $p = \ceil{\log \left(\frac{1 +2\eps}{{\cal O}}\right)}$.} 
   
   Note that $S$ decreases with increase of $\eps$ and hence, the space used by $\cA'$ is $\max(S(\eps_1),\ldots, S(\eps_p))=S(\eps_p)$. Observe that $\eps_p \geq \eps \cdot \cO$. Thus, the space used by $\cA'$ is bounded by $\leq S(\eps \cdot {\cal O})$. 
\end{proof}
 
%\subsubsection{Notations}

\subsection{Our contributions and previous results}
\label{ssec:contrib}

\noindent All lower bounds discussed are randomized lower bounds. In this paper, 
the term {\em hardness} implies 
that without a sketch size $\size{\cS} = \Omega(n)$, we can not solve that problem. The problem statements and results follow.

\subsubsection*{Klee's measure}
\label{sssec:measure-klee}

\begin{itemize}
\item 
(\klee) Given a set of streaming axis-parallel hyperrectangles $\cR=\{R_1,\ldots,R_n\}$ in $\R^d$, the Klee's measure problem is to find the volume $V := \bigcup_{i=1}^{n}R_i$. We show that any $(\eps,\rho)$-additive solution for
Klee's measure requires  a space of $\Omega\left( \frac{1}{\eps} + \log n \right)$ bits even for $d=1$. We give an $(\eps, \rho)$-additive solution for Klee's measure by using $O\left(\frac{1}{\eps^2} \log \left( \frac{1}{\rho} \right) \right)$ space for any constant $d$. For $\delta$-fat\footnote{A hyperrectangle $R \subset [0,1]^d$ is $\delta$-fat if the length of each side is at least $\delta \in (0,1]$. } hyperrectangles, we provide a deterministic algorithm that  uses $O\left( \frac{1}{\eps^{d-1}\delta^d}\right)$ space and gives $\eps$-additive solution to the Klee's measure for constant dimension $d$. 
\end{itemize}

The problem of Klee's measure was first posed by V. Klee~\cite{Klee77} in 1977.
Since then, there have been a series of works done on Klee's
measure~\cite{Bently77,Chan10,Chan13,OvermarsY91,LeewenW81} in the RAM model. The best known algorithmic result in the RAM model, a time complexity of $O(n^{d/2})$, is by Chan~\cite{Chan13}.  We highlight that Klee's measure has connections to estimating $F_0$, the number of distinct elements in a stream. A brief exposition on this is given in Section~\ref{ssec:kleeF0}.
A discrete version of Klee's measure was studied for streaming 
in~\cite{TirthapuraW12} followed by~\cite{SharmaBVRT15}. In both~\cite{TirthapuraW12} 
and~\cite{SharmaBVRT15}, we have a set of points in $\mathbb{Z}^d$ as a discrete universe $U$ and stream $\cR$ of rectangles. The objective is to report the number of points of $U$ present inside $\bigcup_{i=1}^{n}R_i$. Tirthapura and Woodruff~\cite{TirthapuraW12} gave a $(\eps,\rho)$-multiplicative solution to the Klee's measure
 and uses a space of 
$O\left( (\frac{1}{\eps}\log (nU))^{O(1)}\right)$. In~\cite{SharmaBVRT15}, Sharma et al. 
improved the above result for a stream of two dimensional \emph{fat}\footnote{Here {\em fat} rectangle means $\frac{1}{C} \leq \frac{\mbox{width}}{\mbox{height}}\leq C$, for some constant $C >1$.} rectangles using a space of
$O \left(\frac{\log U}{\eps^2}\right)$ bits. They also gave an algorithm for a stream of arbitrary rectangles that gives output which is at most $\sqrt{\log U}$ times the optimal solution and uses a space of $O (\log^2 U \log \log U)$ bits. Note that the space complexity in~\cite{SharmaBVRT15} is independent of $n$. Also, algorithms by Sharma et al.~\cite{SharmaBVRT15} have better update time per rectangle than that of Tirthapura and Woodruff~\cite{TirthapuraW12}.

We give the first lower bound for approximating \klee in $\R$. Our algorithms for \klee works for the original setting of \klee in $\R^d$ as opposed to the discrete versions in \cite{SharmaBVRT15,TirthapuraW12}. Our randomized algorithm assumes nothing on the input rectangles but the deterministic streaming algorithm assumes a fatness condition.

\subsubsection*{Convex body approximation and geometric query}
\label{sssec:measure-query}

\begin{itemize}%[style=unboxed,leftmargin=0.2cm]
\item 
(\convhull) Let $\cCH(\cP)$ denote the convex hull of $\cP$. We
strengthen the lower bound results by showing the promise version of convex hull to be hard, i.e., it is hard to distinguish between inputs having $\size{\cCH(\cP)}=O(1)$ and
$\size{\cCH(\cP)}=\Omega(n)$ in $\R^2$. 

Problems of convex body approximation 
requires storing the approximated convex body. 
We show a space lower bound of $\Omega \left( \frac{1}{\sqrt{\eps}}\right)$ bits for an $(\eps, \rho)$-approximate solution to convex body approximation in $\R^2$. To the best of our knowledge, this is the first data structure lower bound for convex body approximation.
We design an $\eps$-approximate solution for convex body approximation using a space of 
$O(\log^d n / \eps^{\frac{d-1}{2}})$ for a fixed dimension $d$, when it is given to us as a stream of hyperplanes  This implies an $\eps$-additive  
one-pass deterministic streaming algorithm for low dimensional \lp. 
\remove{
\item 
(\coordmax) Let $\cP=\{p_i=(x_i^1, \ldots, x_i^d): 1 \leq i \leq
n\}$ be a set of points in $\R^d$. A point $(x^1,\ldots, x^d) \in
\cP$ is said to be a {\em skyline point} if and only if for each $ i \in [n]$, there exists a $j \in [d]$, such that $x^j \geq x_i^j$. The problem is to
compute the number of skyline points in $\cP$. We show
different kinds of hardness results. 
}
\item 
(\geoqpoly) The decision problem is to detect if a convex polyhedron $\cC$,
 given as an input stream of at most $n$ hyperplanes in $\R^d$ contains a query point $q$ given at the end. We show that this problem is hard
and then obtain a {\em property testing} type result using the
convex body approximation 
result. \remove{We also consider a variant, named \geoqchain, where we 
consider a {\em convex chain} instead of a closed convex polygon.}

In~\cite{ChanC07}, Chan et al.~gave multi-pass algorithms
for computing \emph{exact} convex hull in $\R^{2}$ and $\R^{3}$ along with nearly matching lower bounds for some special class of deterministic algorithms in case of $\R^2$, which was later generalized by Guha et al.~\cite{GuhaM08}. Zhang~\cite{zhangthesis} showed a randomized lower bound of $\Omega(n)$ bits of space for 
the \emph{$k$-promise convex hull} problem
where one knows beforehand that the convex hull has $k$ points. 
We strengthen the lower bound results of the \emph{promise} version of convex
hull given in~\cite{zhangthesis} by showing that it is hard to
distinguish between inputs having $\size{\cCH(\cP)}=O(1)$ and
$\size{\cCH(\cP)}=\Omega(n)$ in $\R^2$.

 \remove{We next prove a randomized lower bound of $\Omega\left(\frac{1}{\sqrt{\eps}}\right)$ bits for $\eps$-approximation of a convex body in $\R^2$. Note that this is the first data structure lower bound for convex body approximation even in the RAM model. 
We also deduce lower bounds for \geoqpoly. We design an algorithm to
approximate a convex body by using ideas from~\cite{AgarwalHV04,BentleyS80,Dudley74}. This helps in approximating low dimensional \lp and designing a property testing result for \geoqpoly for one-pass streaming model.}
\remove{
\item 
(Problem \simpc) Given $n$ points $\cP$ in $\R^2$ in an order (say,
anti-clockwise), the problem is to detect if the points of $\cP$ forms a simple
curve. This is shown to be a hard problem.

\item 
(Problem \klee) Given $n$ intervals (hyper-rectangles) in $\R$ ($\R^d$), the
problem is to find the length (volume) of the union of intervals
(hyper-rectangles). We show this is also hard and then design
approximation results for it.  
}
\end{itemize}

\subsubsection*{Discrepancy}

\begin{itemize}
\item 
(\gdisc) Given  $n$ points $\cP$ as a stream,
where each $p_i \in [0,1]$, the objective is to report $\geo(\cP)$,
the 1-dimensional \emph{geometric discrepancy}~\cite{bookkuipers} of
$\cP$, defined as $$\geo(\cP):=\sup\limits_{[p,q]\subseteq [0,1]}\size{\size{[p,q]} -\frac{n_{pq}}{n}},$$
 where \remove{$l_{pq}=\size{q-p}$ and} $n_{pq}$ is the number of points in $[p,q]$. We show that any $(\eps, \rho)$-additive solution to $\geo(\cP)$ 
requires space bound of $\Omega\left (\frac{1}{\eps} \right)$ bits. We 
present a matching $\eps$-additive deterministic algorithm.

\item 
(\cdisc) Given  $n$ points $\cP$ as a stream,
where each $p_i \in [0,1]$, and a color label red or blue on each 
point, the objective is to report 1-dimensional
\emph{color discrepancy}~\cite{bookmatousek}
 of $\cP$ denoted and defined as $$\col (\cP) =\sup\limits_{I \subseteq [0,1]}|R(I)-B(I)|,$$ where $R(I)$ and $B(I)$
denote the number of red and blue points of $\cP$ respectively, 
that belong to the interval
$I$. We show that any $(\eps, \rho)$-multiplicative solution to $\col(\cP)$ 
admits a space lower bound of $\Omega\left( n \right)$ bits, where $0 < \eps < \frac{1}{5}$. If $\cP$ arrives in a sorted order, $\col(\cP)$ can be computed 
in constant space. \remove{We define a variant of this problem where the interval $I$ is at most a fixed width $\alpha$. It is denoted and defined as $\alphacol(\cP)=\max\limits_{I\subset[0,1], \size{I}\leq \alpha}\size{R(I)-B(I)}$. We show that it is hard to compute $\alphacol(\cP)$ even if $\cP$ arrives in the sorted order and give a $2$-factor deterministic approximation algorithm to compute $\alphacol(\cP)$ for any sorted stream $\cP$.}
\end{itemize}

The only work prior to ours considering discrepancy in the streaming model has been the work by Agarwal et al.~\cite{AgarwalMPVZ06}. They defined \emph{discrepancy} in the context of spatial scan statistics and gave lower bounds and algorithmic results with respect to that. We stick to the conventional definition of both geometric and combinatorial discrepancy and our lower bound results are stronger than that of Agarwal et al.~\cite{AgarwalMPVZ06}. We also remark that $(\eps,\rho)$-additive solution to \gdisc and $(\eps n,\rho)$-additive solution to \cdisc can be found using the algorithm for \emph{all quantile estimation}~\cite{KarninLL16}; the space required for both the cases is $O\left(\frac{1}{\eps} \log \log \frac{1}{\eps} \right)$. Note that the bound achieved by our algorithm for \gdisc is $O\left(\frac{1}{\eps}\right)$ and we are seeking an $(\eps,\rho)$-multiplicative solution for \cdisc.

\remove{In all the above problems except \cdisc, the optimal solutions lie in $[0,1]$ 
and we provide lower bound results for $\eps$-additive solution. 
Note that in such cases, 
the lower bound results for $\eps$-additive solutions are relatively stronger than their $\eps$-multiplicative counterparts.} 

\subsection{A brief review}
\label{ssec:review}
\noindent
Historically, the works of Morris~\cite{Morris78}, Munro and
Paterson~\cite{MP78} and Flajolet et al.~\cite{Flajolet85} were precursors to
the work of Alon et al.~\cite{AlonM99} where the idea of lower bounds on space
for approximating frequency moments was considered. Lower bounds in
streaming is an active area of interest since
then~\cite{HenzingerRR98,Muthu05,Tim16}.  

In computational geometry, researchers have looked at one-pass streaming
algorithms for fundamental problems like convex
hull~\cite{HershbergerS08,RufaiR15}, minimum empty
circle~\cite{AgarwalHV04,AgarwalS15,ChanP14,Zarrabi-ZadehC06}, 
diameter of a point
set~\cite{AgarwalS15,AryaC14,ChanCoreset06,FeigenKZ04,Indyk03soda},
other {\em extent measures} like width, annulus, bounding box, cylindrical
shell~\cite{AgarwalHV04,AndoniN16talg,ChanCoreset06}, 
clustering~\cite{Har-PeledM04}, deterministic $\eps$-net
and $\eps$-approximations~\cite{BagchiCEG07} and their randomized
versions~\cite{FrahlingIS08}, robust
statistics on geometric data~\cite{BagchiCEG07}, geometric
queries~\cite{BagchiCEG07,BishnuCNS15,SuriTZ06}, interval
geometry~\cite{CabelloP15,EmekHR12}, minimum weight matching, $k$-medians and
Euclidean minimum spanning tree weight~\cite{FrahlingIS08,Indyk04stoc}. As
mentioned in~\cite{Indyk04fsttcs}, most of the algorithms for these problems
follow either the \emph{merge and reduce}
technique~\cite{AgarwalHV04,AgarwalS15,AryaC14,ChanCoreset06} or \emph{low
distortion randomized embeddings}~\cite{FrahlingIS08,Indyk03soda,Indyk04stoc}.
In two recent works, \emph{polynomial methods} have been used to find the
approximate width of a streaming point set~\cite{AndoniN16talg} and
$\eps$-kernels~\cite{Chan16} for dynamic streaming (streaming with both
insertion and deletion). On the line of merge and reduce technique, the seminal
work of Agarwal et al.~\cite{AgarwalHV04} on $\eps$-kernels, that is a coreset
for extent measure kind of problems, led to a series of works based on
coresets in
streaming~\cite{AgarwalS15,AryaC14,ChanCoreset06,Chan09dcg,ChanP14,Zarrabi-Zadeh11}. To avoid repetition, we refer the reader to~\cite{Chan16} for a
nice summary on this line of work. 

Streaming algorithm for convex hull in
$\R^2$ was considered in~\cite{HershbergerS08} where by storing $2r+1$ points
one can obtain a distance error of $O(D/r^2)$ ($D$ is the diameter of the point
set) between the original and the reported convex hull. Another
streaming algorithm with an error bound on area of the convex hull of the given points  was proposed in~\cite{RufaiR15}.
Multi-pass streaming algorithm, as the name suggests, can do more than 
one pass on the data stream. Convex hull, linear programming (\lp)
~\cite{ChanC07,GuhaM08} and skyline~\cite{SarmaLNX09} problems have been studied
under the multi-pass model.

Compared to algorithms for geometric problems in streaming, there has not been 
much study on lower bounds in streaming. There are mostly two types of 
lower bound results -- the usual space lower bound of
streaming~\cite{FeigenKZ04,SarmaLNX09}
and the trade off between approximation ratio
and space~\cite{AgarwalS15,BagchiCEG07,Zarrabi-ZadehC06}. 
Feigenbaum et al.~\cite{FeigenKZ04} show that any exact algorithm for computing
the diameter 
of a set of points requires $\Omega(n)$ bits of space.  Zadeh and
Chan~\cite{Zarrabi-ZadehC06} proved a lower bound of $(1+\sqrt{2})/2$ on the 
approximation factor of any deterministic algorithm for the minimum enclosing 
ball (MEB) that at any time stores only one enclosing ball.
Agarwal and Sharathkumar~\cite{AgarwalS15} deduce
lower bounds using the communication complexity model~\cite{Nishan97,Tim16} by
defining $\alpha$-approximate variants 
of MEB, diameter, coreset and width. $\alpha > 1$ is 
a multiplicative parameter on the radius of MEB, the MEB of 
the coreset, the diameter and the width of the slab containing the points. Defining this    
$\alpha$-approximate variants, allow them to deduce lower bounds on space 
in terms of bits in the communication complexity model and relate approximation
bounds to the space by obtaining suitable values of $\alpha$. All of their 
lower bound results are in the following framework -- any streaming algorithm 
that maintains an $\alpha$-MEB, an $\alpha$-diameter, an $\alpha$-coreset or 
an $\alpha$-width for a set of points in $\R^d$, for $\alpha < C$ (where $C$
is a constant depending on $d$ and is different for each of the problems), with
probability at least $2/3$ requires $\Omega( \mbox{min}
\{n,\mbox{exp}(d^{1/3})\})$ bits of storage. Apart from the above, Bagchi et al.~\cite{BagchiCEG07} 
showed that it is not possible to approximate the range counting problem in 
polylogarithmic space. 

\remove{Of relevance to us, 
is the lower bound results on skyline, convex hull and
closest pair~\cite{SarmaLNX09,zhangthesis,ChanC07,GuhaM08}. 
Das Sarma et al.~\cite{SarmaLNX09} deduce a lower bound
on the skyline problem that states that any randomized algorithm can not find
with high probability the $m$ skyline points from an input of $n+m$ points by
storing less than $n/2$ points. As the skyline points are to be reported, the
lower bound is in terms of storing points and not bits that is usual in
streaming lower bounds. In effect, lower bounds in terms of storing points
becomes important for other problems like 
convex hull and \lp. Lower bounds and near matching upper bounds for convex hull
and \lp in the multi-pass streaming model was first considered
in~\cite{ChanC07}. The lower bound results were later on generalized 
in~\cite{GuhaM08}. Their results show that any convex hull algorithm that
receives the stream of points in $\R^2$ sorted in their $x$-coordinates needs
at least $\Omega(\sqrt{n})$ points to be stored. For \lp in fixed dimensions,
the corresponding lower bound on space is $\Omega(n)$ points.
Zhang~\cite{zhangthesis} considered the
\emph{$k$-promise convex hull} problem (these type of problems belong to the
class of \emph{promise problems}) where one knows beforehand that the convex
hull has $k$ points. They proved that any randomized algorithm for 
 $k$-promise convex hull requires $\Omega(n)$ bits of space.
Zhang~\cite{zhangthesis} also proved that any randomized algorithm that solves
closest pair requires $\Omega(n)$ bits of space. 
}
All our lower bounds will be stated in number of bits that is consistent with the 
streaming model, where as the upper bounds will be stated in number of words. The lower bounds will be based on communication complexity arguments by using the results on \ind and \disj problems. In any instance of the 
\ind problem, Alice has ${\bf x} \in \{0,1\}^n$ and Bob has an integer (index) $i$. The goal is to compute the $i$-th bit of $\bf{x}$ i.e., $x_{i}$. We say $\ind ({\bf x}, i)=1$ if and only if $x_i=1$. In an instance of the 
\disj problem, both Alice and Bob have bit vectors ${\bf x, y}\in \{0,1\}^n$. The goal is to determine whether there exists $i \in [n]$ such that $x_i=y_i=1$.  We say $\disj ({\bf x},{\bf y})=0$ if and only if there exists $i \in [n]$ such that $x_i=y_i=1$. \ind is hard in one-way communication complexity and \disj is hard in two-way communication complexity. The following, stated as a Theorem, will be useful for us.
\remove{can be summarized as follows. }

\begin{theo}~\cite{Nishan97,Tim16}\label{theo:index_disj}
\begin{itemize}
\item[(a)] Every randomized one-way protocol that solves \ind problem with
probability at least $1-\rho$ uses $\Omega(n)$ bits of communication. 
\item[(b)] Every randomized two-way protocol that solves \disj problem with
probability at least $1-\rho$ uses $\Omega(n)$ bits of communication. 
\end{itemize}

\end{theo}

\remove{
 In two way communication
 complexity \disj is hard and formally stated as follows.
\begin{theo}~\cite{Nishan97,Tim16}
\label{theo:disj}
Every randomized two way protocol that solves \disj problem with
probability at least $1-\rho$  uses $\Omega(n)$ bits of communication. 
\end{theo}}
\remove{Both \disj and \ind can used to prove \emph{space lower bound} for one pass streaming algorithm 
for a problem $P$. Let \disj reduces to a problem $P$ such that any one pass streaming
algorithm for $P$ using space $s$ can be used to solve \disj. Then observe that we can have a reduction from \disj to $p$-pass streaming algorithm for $P$ 
If one pass streaming algorithm for a problem $P$, that uses space $s$, can be used to solve \disj, then
$p$-pass }
\remove{
\subsection{Notations}
We will delete this subsection at the end. We will put all notations here 
to have consistency. 
\begin{description}
\item $\R$ -- real 
\item $\cS$ -- sketch 
\item $\cP$ -- the stream (mostly points)
\item $\cCH(\cP)$ -- the convex hull of $\cP$
\item $[n]$ -- $\{1, \ldots, n\}$
\item $\cP=\{p_i = (x_i^1, \ldots, x_i^d)\}$
\item $\cC$ -- convex polygon/polyhedron/chain
\item $\delta(\cdot\, \cdot)$ -- distance
\item $\alpha$ $\in [n]$ is a parameter for proximity query
\item static part of dynamic sketch -- $S$ 
\item dynamic part of dynamic sketch -- $D$
\item diameter -- $D$ 
\end{description}
}

\section{\textsc{KLEE'S MEASURE}}
\label{sec:klee}
This section begins with a discussion that shows the importance of klee's measure in the sense that 
it is related to $F_0$ estimation. Next, we present lower bounds along with randomized and deterministic algorithms.

\subsection{Connection of Klee's measure to $F_0$ estimation}
\label{ssec:kleeF0}

 Let us consider a stream $\cR$, where each $R_i \subseteq [N]^d$, i.e. corners of each rectangle have integer coordinates. Recall that klee's measure of $\bigcup\limits_{i=1}^n R_i$ is denoted as $V$. Our objective is to report the estimate $\hat{V}$ of the volume of $\bigcup\limits_{i=1}^n R_i$ such that 
 $\size{V-\hat{V}}\leq \eps V$. The following result will be of importance.

Let $F_0$ be the number of distinct elements present in a stream such that each element in the stream is from universe $U$. Then, there exists a one pass randomized streaming algorithm
$ALG$ that finds $F_0'$ such that $\size{F_0-F_0'} \leq \eps F_0$ and
uses $O \left(\frac{1}{\eps^2} + \log \size{U} \right)$ bit of space, where 
$\eps$ is an input parameter~\cite{JelaniW10}. This is the optimal algorithm w.r.t. space for $F_0$ estimation~\cite{AlonM99, Woodruff04}.

 Here corners of each $R_i$ have integer cordinates. This implies that each rectangle is a disjoint union of unit hypercubes that lies inside it. So, Klee's measure $V$ is the number of distinct unit hypercubes in $\bigcup_{i=1}^{n}R_i$. On receiving a hyperrectangle $R_i$ in the stream, 
 we give all the unit cubes inside $R_i$ as inputs to $ALG$. At the end of the stream,
 we report the ouput produced by $ALG$ as $\hat{V}$. Observe that $\size{V-\hat{V}}\leq \eps V$. Note that here the size of the universe for $F_0$ estimation is $[N]^d$. Hence, we have the following observation.
 \begin{obs}
\label{theo:klee_dist}
Let a stream of hyperrectangles be such that corners of each rectangle 
have integer coordinates in $[N]^d$. Then, there exists a randomized one-pass streaming algorithm that outputs $\hat{V}$ such that $\size{V-\hat{V}}\leq \eps V$ with high probability and uses $O\left(\frac{1}{\eps^2} + d\log N \right)$ bits of space. 
\end{obs}
One can reduce $F_0$ estimation to the Klee's
measure problem where corners of each rectangle have integer coordinates as follows. So, the space used by the corresponding algorithm of Observation~\ref{theo:klee_dist} is optimal.

\subsection{Lower bound}
\label{ssec:klee_lb}

\remove{ In the following Section, $\cI(a,b)$ denotes an interval $[a,b] \in \R$.} 
\begin{figure}[!h]
%\begin{minipage}{0.5\textwidth}
\centering
\includegraphics[width=0.8\linewidth]{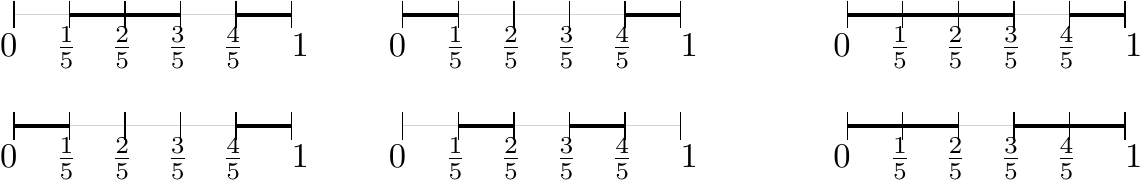}
  \caption{Reduction idea for Theorem~\ref{theo:klee_lb}.
  Here $n=5$. In both (a) and (b), Alice's input ${\bf x}$ is shown in the left figure and Bobs input ${\bf y}$ is shown in the middle one. In (a), ${\bf x}=01101$ and ${\bf y}=10001$; $\disj({\bf x},{\bf y})=0$; $l_A=\frac{3}{5}$ and $l_B=\frac{3}{5}$; the total Klee's measure is shown in the right figure and the value is $\frac{4}{5} < l_A+l_B$. In (b), ${\bf x}=10001$ and ${\bf y}=01010$; $\disj({\bf x},{\bf y})=1$; $l_A=\frac{2}{5}$ and $l_B=\frac{2}{5}$; The total Klee's measure is shown in the right figure and the value is $\frac{4}{5} = l_A+l_B$.}
  \label{fig:klee_lb}
\end{figure}
\begin{theo}
\label{theo:klee_lb}
For every $\eps \in (0,0.1)$, there exists an positive integer $n$ such that any randomized one-pass streaming algorithm that outputs $(\eps, \rho)$-additive solution to \klee for all streams of at most $2n$ intervals in $\R$, uses $\Omega \left( \frac{1}{\eps}+ \log n \right)$ bits of space.
\end{theo}

\begin{proof}
We prove the Theorem by proving the followings.
\begin{description}
\item[(a)] For every $\eps \in (0,0.1)$, there exists an positive integer $n$ such that any randomized one-pass streaming algorithm that outputs $(\eps, \rho)$-additive solution to \klee for all streams of at most $2n$ intervals in $\R$, uses $\Omega \left( \frac{1}{\eps} \right)$ bits of space.
\item[(b)] For every $\eps \in (0,0.1)$, there exists an positive integer $n$ such that any randomized one-pass streaming algorithm that outputs $(\eps, \rho)$-additive solution to \klee for all streams of at most $2n$ intervals in $\R$, uses $\Omega \left( \log n \right)$ bits of space.
\end{description} 

(a)\remove{ Without loss of generality, assume that $0 < \eps < \frac{1}{4}$. Otherwise the Theorem trivially follows.} Let $n=\floor{\frac{1}{2\eps}}-1$ and ${\cal A}$  be a streaming algorithm that returns $(\eps, \rho)$-additive  solution to $\klee$ and uses space $s=o\left(\frac{1}{\eps}\right)$. The following one-way communication protocol can solve the \disj using space $o(n)$. Alice will process her input ${\bf x}$ as follows. See Figure~\ref{klee_lb}. For each $i$, if $x_i=1$ then we give the interval $[\frac{i-1}{n}, \frac{i}{n}]$ as input to $\cA$, otherwise, we do nothing. Let the corresponding set of 
intervals generated by Alice be $S_A$ and $l_A$ be the Klee's measure of $S_A$. Alice will send the current memory state of $\cA$ and $l_A$ to Bob. Similarly, Bob will process his input $\bf y$.
Let the corresponding set of 
intervals generated by Bob be $S_B$ and $l_B$ be its Klee's measure. Observe that $\disj({\bf x, y})=1$ if the length of $S_A \cup S_B$ is $l_A+  l_B$ and in this case $\cA$ returns at least $l_A+l_B-\eps$. $\disj({\bf x, y})=0$ if the length of $S_A \cup S_B$ is at most $l_A + l_B -\frac{1}{n}<l_A + l_B -2\eps$ and  $\cA$ returns less then $l_A + l_B - \eps$. So, we report $\disj({\bf x, y})=1$ if and only if $\cA$ gives output at least $l_A+l_B-\eps$. 
 \remove{Now give input
$\cI(\frac{j-1}{n},\frac{j}{n})$ to $\cA$. Note that if $x_j=1$, the
input interval given at the end does not add anything to $l$, otherwise it
adds $\frac{1}{n}$ to $l$.Observe that $\disj({\bf x, y})=1$ if and only if the output of $\cA$ is $l_A+l_B$.}

(b) Let $\cV$ be a family of $\{0,1\}^n$ vectors such that $\forall{\bf x} \in \cV$, $\sum\limits_{i=1}^{n}x_i=\frac{n}{2}$; $\forall{\bf x}, {\bf y} \in \cV$ and ${\bf x} \neq {\bf y}$ ,  there are at most $\frac{n}{4}$ indices where both ${\bf x}$ and ${\bf y}$ have $1$.\remove{${\bf x}.{\bf y} \leq \frac{n}{4}$. It can be easily shown the existence of such a family $\cV$, where $\size{\cV}=2^{\Omega(n)}$.} Such a family $\cV$ with $\size{\cV}=2^{\Omega(n)}$ exists~\cite{AlonM99}. We define the \eql function as follows. Both Alice and Bob get vectors ${\bf x}$ and ${\bf y}$ respectively from $\cV$ and the objective is to decide whether ${\bf x}={\bf y}$. Formally, $\eql({\bf x}, {\bf y})=1$ if and only if ${\bf x}={\bf y}$. It is well known that the two-way private coin randomized communication complexity of \eql is $\Omega(\log n)$~\cite{AlonM99, Tim16, Nishan97}.

\remove{With out loss of generality assume that $0 < \eps < \frac{1}{10}$.} Let $ n > 2^{1/\eps}$ and ${\cal A}$ be a streaming algorithm that returns $(\eps, \rho)$-additive solution to $\klee$ using space $o(\log n)$ bits. The following one-way communication protocol can solve the \eql using space $o(\log n)$. Both Alice and Bob process their input exactly as in part (a). Let $S_A$, $S_B$, $l_A$, $l_B$ have the same notation as in (a). Observe that if $\eql({\bf x},{\bf y})=1$, then the length of $S_A \cup S_B$ is $0.5$ and in this case $\cA$ returns at most $0.5+0.1=0.6$. If $\eql({\bf x},{\bf y})=0$, then the length of $S_A \cup S_B$ is at least $0.75$ and in this case $\cA$ returns at least $0.75-0.1=0.65$. Therefore $\eql({\bf x, y})=1$ if and only if $\cA$ gives output at most $0.6$. 
\end{proof}
\begin{rem}
\label{rem:klee_lb}
\begin{itemize}
\item Multipass lower bound: We can also do both of the above reductions from \disj and \eql to a $p$-pass 
streaming algorithm, using space $s$, for \klee in $\R$. Observe that the induced protocol for \disj requires at most $2ps$ bits of communication. This implies $s =\Omega\left ( \frac{1}{p} \left( \frac{1}{\eps}+ \log n\right)\right)$.
%\item Our discussion in Section~\ref{ssec:kleeF0} says that any one-pass streaming algorithm that returns $(\eps, \rho)$-multiplicative solution to \klee, requires $ \Omega \left( \frac{1}{\eps^2} +\log n \right)$ bits of space. 
\end{itemize}
\remove{
\begin{itemize}
\item  Multipass lower bound: We can also do the above reduction from \disj to a $p-$pass 
streaming algorithm, using space $s$, for \klee in $\R$. Observe that the induced protocol for \disj requires at most $2ps$ bits of communication. This implies $s =\Omega\left ( \frac{n}{p}\right)$. 
\item Stronger one pass lower bound: One can also reduce $F_0$ estimation to a variation of klee's
measure problem, where end points of all the intervals are integer coordinates and each interval are
of unit length. So, the lower bound for $F_0$ estimation holds for one pass lower bound for \klee.
\end{itemize}
}
\end{rem}

\subsection{Algorithms for Klee's measure in $[0,1]^d$ }

\subsubsection*{Randomized algorithm}

 Let us consider a stream $\cR$, where each $R_i \subset [0,1]^d$. Our objective 
 is to output $\hat{V}$ such that $\size{V-\hat{V}}\leq \eps$ with probability at 
 least $1-\rho$, where $0 < \eps, \rho < 1$. \remove{A random sampling based idea gives the required output
 using $O\left( \frac{1}{\eps^2} {\log \left(\frac{1}{\rho}\right)}\right)$ space.}
We generate and store $M$ random points $p_1,\ldots, p_M \in [0,1]^d$. For each point $p_j, j \in [M]$, we maintain a binary indicator random variable $X_j$, $j\in [M]$. $X_j=1$ if and only if $p_j$ lies on or inside any $R_i \in \cR$. \remove{ 
We process each rectangle $R_i$, $i \in [n]$, in the stream by setting $X_j=1$ if and only if $p_j \in R_i$, $j \in [M]$.} At the end of the stream, we report $\hat{V}=\frac{X}{M}$, where $X=\sum_{i=1}^{M} X_i$.

As we have chosen $M$ points uniformly at random, the probability that a random point is present in some 
rectangle in the stream, is same as the volume of $\bigcup\limits_{i=1}^n R_i$. So, $\p(X_j=1)=V$ and $\mathbb{E}({X})=MV$. Now apply
standard Chernoff bound. 
\begin{eqnarray*}
\p\left(\size{V-\hat{V}} \geq \eps \right) &=& \p(\size{X-MV} \geq M\eps) \\
 &\leq& 2e^{-\frac{(\eps/V)^2}{2+\eps/V} MV} \\
  &=& 2e^{-\frac{\eps^2}{2 V+\eps} M} \\
   &\leq& 2e^{-\frac{\eps^2}{2 +\eps} M} \leq \rho.
\end{eqnarray*}
This implies $M \geq \frac{2+\eps}{\eps^2}\log \left( \frac{1}{\rho} \right)$. Hence, we have the following theorem.

\begin{theo}
\label{theo:klee_rand}
There exists a randomized one-pass streaming algorithm that outputs $(\eps,\rho)$-additive solution to Klee's measure in $[0,1]^d$ and uses $O\left(\frac{1}{\eps^2}\log \left(\frac{1}{\rho}\right) \right)$ space. 
\end{theo}

\subsubsection*{Deterministic algorithm}
\label{ssec:klee_det}

A hyperrectangle $R \subset [0,1]^d$ is $\delta$-fat if the length of 
each side is at least $\delta \in (0,1)$.
\remove{Let $\cR = \{R_1,\ldots,R_n\} \subset [0,1]^d$ be a stream of $\delta$-fat hyperrectangles and $V$ be the volume (Klee's measure) of $\bigcup\limits_{i=1}^n R_i$.} Our objective is to report a $\hat{V}$ such that $V-\eps \leq \hat{V} \leq V$ for a given $\eps \in (0,1)$. The main result is stated in the following theorem.

\begin{algorithm}
%\scriptsize
\caption{Klee($d,\eps$)}
\label{algo_klee_main}
\KwIn{A stream of hyperrectangles $\cR=\{R_1,\ldots,R_n\}$ in $[0,1]^d$.}
\KwOut{Klee's measure of the stream}
%\LinesNumbered
\Begin
	{
	Subdivide $[0,1]^d$ into $1/\delta^d$ hyperboxes, each of size $\delta \times \ldots \times \delta$ as shown in Figure~\ref{fig:klee}; denote each such hyperbox as $\cH^{\delta}_d$.\\
	We magnify each dimension of $[0,1]^d$ by $1/\delta$ so that each $\cH^{\delta}_d$ becomes $[0,1]^d$.\\ 
Let $\cB$ be the set of all \emph{magnified} ${\cH}^{\delta}_d$'s.\remove{ Recall that each hyperrectangle $R \in \cR$ is $\delta$-fat. Therefore, for each ${\cH}^{\delta}_d \in \cB$, if $R \cap {\cH}^{\delta}_d \neq \phi$, then $R \cap \cH_{\delta}^d$ is anchored at (at least) one of the corners of ${\cH}^{\delta}_d$. Assign $R$ to one such corner.}\\
 
Call in parallel, $1/\delta^d$ copies of $ALG(d,\eps)$ --- one for each 
$\cH_{\delta}^d \in \cB$. ($ALG(d,\eps)$ is given as Algorithm~\ref{algo_klee_d}.)\\
   \For {(each hyperrectangle $R$ in the stream) }
   {
  Find all $ {\cH_{\delta}^d}\in \cB$ that intersects $R$ and give $R \cap {\cH}_{\delta}^d$ as input to the corresponding $ALG(d,\eps)$ algorithm for ${\cH}^{\delta}_d$.}
	
  Find the sum of outputs produced by all $1/\delta^d$ copies of $ALG(d,\eps)$ and multiply by $\delta^d$ and \emph{return}.
}
\end{algorithm}

\begin{theo}
\label{theo:klee_det}
There exists a deterministic one-pass streaming algorithm that takes a stream of $\delta$-fat hyperrectangles in $[0,1]^d$ as input, and outputs $\hat{V}$ such that $V - \eps \leq \hat{V} \leq V $, using $O\left(\frac{2^{d^2+d}}{\eps^{d-1} \delta^d}\right)$ space, where 
$\eps \in (0,1)$ is an input parameter.
\begin{figure}
  \centering
  \includegraphics[width=1.0\linewidth]{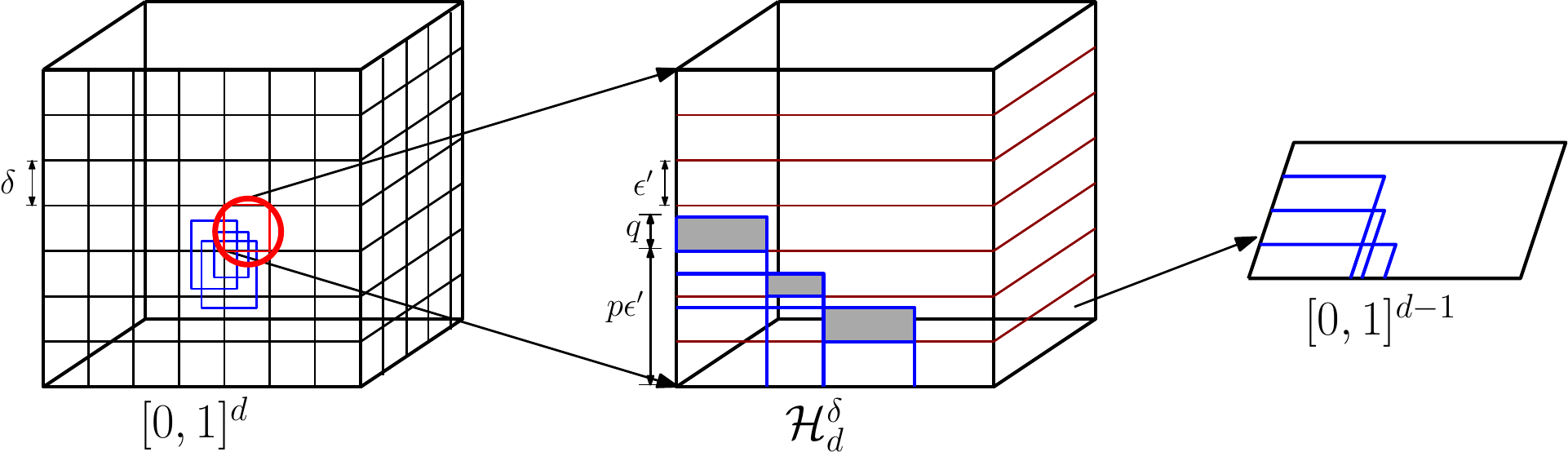}
  \caption{The figure to the left shows the placement of some input rectangles in $[0,1]^d$, the figure in the middle shows anchored rectangle at origin in $[0,1]^d$ and the figure to the right shows projection of all intersection of all rectangles with a particular strip to $[0,1]^{d-1}$.}
  \label{fig:klee}
\end{figure}
\remove{
\item[(b)] There exists a deterministic $\log \left(\frac{1}{V}\right)$-pass streaming algorithm that takes a stream of $\delta$-fat hyperrectangles in $[0,1]^d$ as input, and outputs $\hat{V}$ such that $(1 - \eps)V \leq \hat{V} \leq V $, using $O\left(\frac{2^{d^2+d}}{(\eps V)^{d-1} \delta^d}\right)$ space, where 
$\eps \in (0,1)$ is an input parameter.
}
\end{theo}

\begin{proof}
 The crux of the proof is to subdivide $[0,1]^d$ into $1/\delta^d$ hyperboxes, each of size $\delta \times \ldots \times \delta$ as shown in Figure~\ref{fig:klee}; denote each such hyperbox as $\cH^{\delta}_d$. Each $\delta$-fat hyperrectangle of the stream will intersect at least one corner of some 
$\cH^{\delta}_d$; there are $2^d$ such corners for each ${\cH}^{\delta}_d$. For any corner, a subset of hyperrectangles of $\cR$ will be called \emph{anchored} if each member of the subset intersects that corner point.\remove{ Our goal is to approximate $V$ by solving Klee's measure for hyperrectangles of $\cR$ anchored at each of the $O(2^d/\delta^d)$ corners.} A set of hyperrectangles is \emph{anchored} for a hyperbox if each hyperrectangle in the set intersects at least one corner of the hyperbox.  
The next Claim, proved later, finds the estimate of klee's measure for hyperrectangles that are anchored for a hyperbox.

\begin{cl}
\label{clm:klee_anch}
Let ${\cR}_a$ be a stream of hyperrectangles anchored for $[0,1]^d$. There exists a deterministic one-pass streaming algorithm that outputs $\hat{V_a}$ such that 
$V_a-\eps \leq \hat{V_a} \leq V_a$ and uses $O\left(\frac{2^{d^2+d}}{\eps^{d-1}}\right)$ space, where $V_a$ is the Klee's measure of ${\cR}_a$ and $\eps \in (0,1)$ is an input parameter.
\end{cl}

To put the proof of Theorem~\ref{theo:klee_det} in Claim~\ref{clm:klee_anch}'s context of $[0,1]^d$, we magnify each dimension of $[0,1]^d$ by $1/\delta$ so that each $\cH^{\delta}_d$ becomes $[0,1]^d$. The error term will also change accordingly. So, our problem boils down to finding the Klee's measure in $[0,1/\delta]^d$ within an additive error of $\eps/\delta^d$. 
Let $\cB$ be the set of all \emph{magnified} ${\cH}^{\delta}_d$. Recall that each hyperrectangle $R \in \cR$ is $\delta$-fat. Therefore, for each ${\cH}^{\delta}_d \in \cB$, if $R \cap {\cH}^{\delta}_d \neq \phi$, then $R \cap \cH_{\delta}^d$ is anchored at (at least) one of the corners of ${\cH}^{\delta}_d$.\remove{ Assign $R$ to one such corner.}
 
We start, in parallel, $1/\delta^d$ $d$-dimensional anchored Klee's measure algorithm -- one for each 
$\cH_{\delta}^d \in \cB$. On receiving a rectangle $R$, we find all $ {\cH_{\delta}^d}\in \cB$ that intersects $R$ and give $R \cap {\cH}_{\delta}^d$ as input to the corresponding
algorithm for ${\cH}^{\delta}_d$. Refer the pseudocode given in Algorithm~\ref{algo_klee_main}. Total additive error of $\eps/\delta^d$ can be achieved if we can ensure additive error of at most $\eps$ for each $ {\cH}^{\delta}_d \in \cB$. By Claim~\ref{clm:klee_anch}, this can be achieved by using $O\left(\frac{2^{d^2+d}}{\eps^{d-1}}\right)$ space for each ${\cH}^{\delta}_d$. Hence, the total amount space required is $O\left(\frac{2^{d^2+d}}{\eps^{d-1} \delta^d}\right)$.
\remove{
\comments{At the end of the stream, we find the sum of outputs of all $1/{\delta}^d$ copies of $d$-dimensional anchored Klee's measure algorithm, then multiply by $\delta ^d$ and return it as the estimate of Klee's measure.}}
\end{proof}

\begin{proof}[Proof of Claim~\ref{clm:klee_anch}]
We will prove it by using induction on $d$, where $d \geq 1$. Let $ALG(\eps,d)$ be the algorithm that solves the problem within an additive error of $E(\eps, d)$ using $S(\eps, d)$ space.
We have to prove that 
$E(\eps,d)\leq \eps$  and $S(\eps,d) \leq \frac{2^{d^2+d}}{\eps^{d-1}}, ~\mbox{where}~ d\geq 1$. 
For the base case of $d=1$, each interval of the stream in $[0,1]$ is anchored either at 0 or 1. Our algorithm $ALG(\eps,1)$ maintains the rightmost (leftmost) extreme point {\em left} ({\em right}) of all intervals anchored at 0 (1). At the end of the stream, we report $\min(\mbox{{\em left}} + \mbox{{\em right}},1)$. Observe that we output exact Klee's measure without any error using constant space. Hence, $ E(\eps,1)=0\leq \eps ~\mbox{and}~ S(\eps,1)=O(1)$.

\remove{
is as follows. We maintain two variables {\em left} and {\em right} to maintain maximum interval covered anchored at $0$ and $1$, respectively. On receiving an interval $\cI$, we will set {\em left} $ = \max(\mbox{{\em left}},|\cI|)$ or {\em right} $= \max(\mbox{{\em right}},|\cI|)$ depending on whether $\cI$ is anchored at $0$ or $1$. At the end of the stream, we report $\min(\mbox{{\em left}} + \mbox{{\em right}},1)$. Observe that our reported output is exact Klee's measure without any error for anchored intervals in $[0,1]$ using constant space. Hence, $ E(\eps,1)=0\leq \eps ~\mbox{and}~ S(\eps,1)=O(1)$. 
}
Assuming the statement is true for all dimensions less than or equal to $d-1$, we show that it is also true for dimension $d$. 
We divide the entire space $[0,1]^{d}$ along the $d$-th dimension into $1/\eps'$ strips, each of the form  $[0, 1]^{d-1} \times [(i-1)\eps',i\eps']$, where $\eps'=\eps/2^{d+1}$ and $1 \leq i \leq 1/\eps'$. We start $1/\eps'$ copies of the algorithm $ALG(\eps/2,d-1)$, one for each strip.

\remove{
We divide the entire space $[0,1]^{d}$ in to $1/\eps'$ strips along $d^{th}$ dimension, where $\eps'=\eps/2^{d+1}$. Each strip is of the form $[0, 1]^{d-1} \times [(i-1)\eps',i\eps']$, where $0 \leq i \leq 1/\eps'$. We start $1/\eps'$ copies of the algorithm $ALG(\eps/2,d-1)$, one for each strip.}

On receiving an anchored hyperrectangle $R\subset [0,1]^{d}$, we assign it to one of the $2^{d}$ corners with which it intersects. Without loss of generality, assume that it is anchored at the origin. Let $R$ cut through $p$ strips $(0 \leq p \leq 1/\eps')$ and extend for a distance of $q$ $(0 \leq q < \eps')$ in the last strip along the $d$-th dimension. Refer Figure~\ref{fig:klee}. Thus $R$ can be decomposed as $R = R' \times [0,p\eps'+ q]$, where $R' \mbox{ (of dimension $d-1$) } \subset [0,1]^{d-1}$. Divide $R$ into $p$ parts of the form $R'\times [(j-1)\eps', j\eps']$, where $j \in [p]$ and assign $R'\times [(j-1)\eps', j\eps']$  to the $j$-th strip. Part of hyperrectangles we are assigning to a strip is of length $\eps'$ along dimension $d$. So, each anchored hyperrectangle in $[0,1]^d$ assigned to a strip can be thought of as a $d-1$ dimensional hyperrectangle with a length of $\eps'$ along the $d$-th dimension. Observe that the projection of hyperrectangles of the form $R' \times [(j-1)\eps', j\eps']$,  
that belong to a particular strip $j$, to the $d-1$ dimensional space is nothing but $R'$. See Figure~\ref{fig:klee} for an example. For each $j \in [p]$, $R'$ is given as an input to the coresponding $ALG(\eps/2, d-1)$ of the $j^{th}$ strip.\remove{We make a recursive call to $ALG$ with $R'$ as an input and error parameter $\eps/2$ and the dimension as $d-1$, i.e., $ALG(\eps/2,d-1)$.} At the end of the stream, we compute the sum of outputs produced by all $\frac{1}{\eps'}$ recursive calls $ALG(\eps/2,d-1)$ and then multiply by $\eps'$ to get the final estimate of Klee's measure\remove{ \comments{of anchored rectangles}}. Refer the pseudocode given in Algorithm~\ref{algo_klee_d}.

\begin{algorithm}
%\scriptsize
\caption{ALG($d,\eps$): $d$ dimensional anchored Klee's measure, $d > 1$}
\label{algo_klee_d}
\KwIn{A stream of hyperrectangles in $[0,1]^d$ anchored either at $0$ or $1$.}
\KwOut{Klee's measure of the stream}
%\LinesNumbered
\Begin
	{
	 Divide the entire space $[0,1]^{d}$ along the $d$-th dimension into $1/\eps'$ strips, where $\eps'=\eps/2^{d+1}$, such that each strip is of the form $[0, 1]^{d-1} \times [(i-1)\eps',i\eps']$, $1 \leq i \leq 1/\eps'$.\\
	 We start $1/\eps'$ copies of the algorithm $\mbox{ALG}(d-1,\eps/2)$, one for each strip.\\
	\For {(each anchored hyperrectangle $R$ in the stream) }
	{
	Assign $R$ to one of the $2^{d}$ corners with which it intersects. W.l.o.g., 
	assume that it is anchored at the origin.\\
	 Let $R=R'\times[0,p\eps'+ q]$, where $R' \subset [0,1]^{d-1}$, $0 \leq p \leq 1/\eps'$ and 
	 $0 \leq q < \eps'$. Divide $R$ into $p$ parts of the form $R'\times [(j-1)\eps', j\eps']$, 
	 where $j \in [p]$ and assign $R'\times [(j-1)\eps', j\eps']$  to the $j$-th strip. Ignore
	  remaining portion of $R$, i.e, $R'\times [p\eps',p\eps'+ q]$ (the shaded part in Figure~\ref{fig:klee}).\\
	 For each $j\in [p]$, give input $R'$ to the corresponding $\mbox{ALG}(d-1,\eps/2)$ of the $j$-th strip.\\
	 	}
   Find the sum of outputs produced by all $1/\eps'$ recursive calls $\mbox{ALG}(d-1,\eps/2)$ and then multiply by $\eps'$ and \emph{return}.
}
\end{algorithm}

Our algorithm incurs two types of errors -- 
\begin{description}
\item[(i)] Error incurred from the recursive calls to the lower dimensions.
\item[(ii)] The error incurred at this level of the recursion corresponding to the top part of $R$, i.e., $ R' \times [p\eps',p\eps'+ q]$ (as shown 
by the shaded parts in Figure~\ref{fig:klee}) that we just ignore.
\end{description}
  To analyze the error of type (ii), notice that these top parts of the hyperrectangles anchored at a corner has a special structure as these hyperrectangles form a partial order under inclusion. So, the errors are additive and can be at most $\eps'$. Hence, the total error with respect to all $2^d$ corners is at most $2^d \eps'$. Now to estimate the error of type (i), we observe that the error in calculation of Klee's measure due to one strip (because of the multiplication by $\eps'$) is $\eps' E(\eps/2,d-1)$. So, the total error with respect to all the $1/\eps'$ strips is $\frac{1}{\eps'}(\eps'E(\eps,d-1)) = E(\eps,d-1)$ which by induction hypothesis is at most $\eps/2$. Now considering the fact that $\eps'=\eps/2^{d+1}$, the total error is given by $$E(\eps, d) = 2^{d}\eps' + E(\eps/2, d-1) \leq \eps$$. 

With the space requirement for one strip being $S(\eps/2, d-1)$ by induction hypothesis, the total space requirement with respect to all $1/\eps'$ strips is $\frac{1}{\eps'}S(\eps/2,d-1)$ along with the book keeping of
 $\frac{1}{\eps'}$ instances of $ALG(\eps/2,d-1)$.  Therefore, $$S(\eps,d) \leq \frac{1}{\eps'}S(\eps/2,d-1)+ \log \left( \frac{1}{\eps'}\right) =\frac{2^{d+1}}{\eps}S(\eps/2, d-1) + \log \left( \frac{2^{d+1}}{\eps'}\right) \leq \frac{2^{d^2 + d}}{\eps^{d-1}}.$$
\end{proof}

\section{\textsc{CONVEX-BODY} and \textsc{GEOM-QUERY-CONV-POLY}}
\label{sec:extent}

In this section, we first strengthen the lower bound result on convex hull by showing  that even promised 
version in hand. Next, we derive lower bounds for convex body approximation and point existence queries in a convex polygon followed by algorithms for them. To the best of our knowledge, the lower bound on convex body approximation is first of this kind. As a consequence of our result on convex body approximation, we can solve \lp in the streaming model and design a property testing type result for geometric queries.

\subsection{Lower bounds}
\label{ssec:convbody_lb}

\remove{\label{ssec:convhull}}
%We strengthen the lower bound results of convex hull given
%in~\cite{zhangthesis} by showing that it is hard to
%distinguish between inputs having $\size{\cCH(\cP)}=O(1)$ and
%$\size{\cCH(\cP)}=\Omega(n)$ in two dimensions. Next we describe a method to
%approximate a convex body by using ideas
%from~\cite{Dudley74,BentleyS80,AgarwalHV04}.

\subsubsection*{Lower bound for promise version of convex hull}

\begin{theo}\label{lem:hardconv}
Any randomized one-pass streaming algorithm for convex hull that distinguishes
between inputs having $\size{\cCH(\cP)}=O(1)$ and $\size{\cCH(\cP)}=\Omega(n)$
in $\R^2$ with probability $1-\rho$, uses $\Omega(n)$ bits of space.
\end{theo}
\remove{
\begin{figure}
  \centering
  \includegraphics[width=0.4\linewidth]{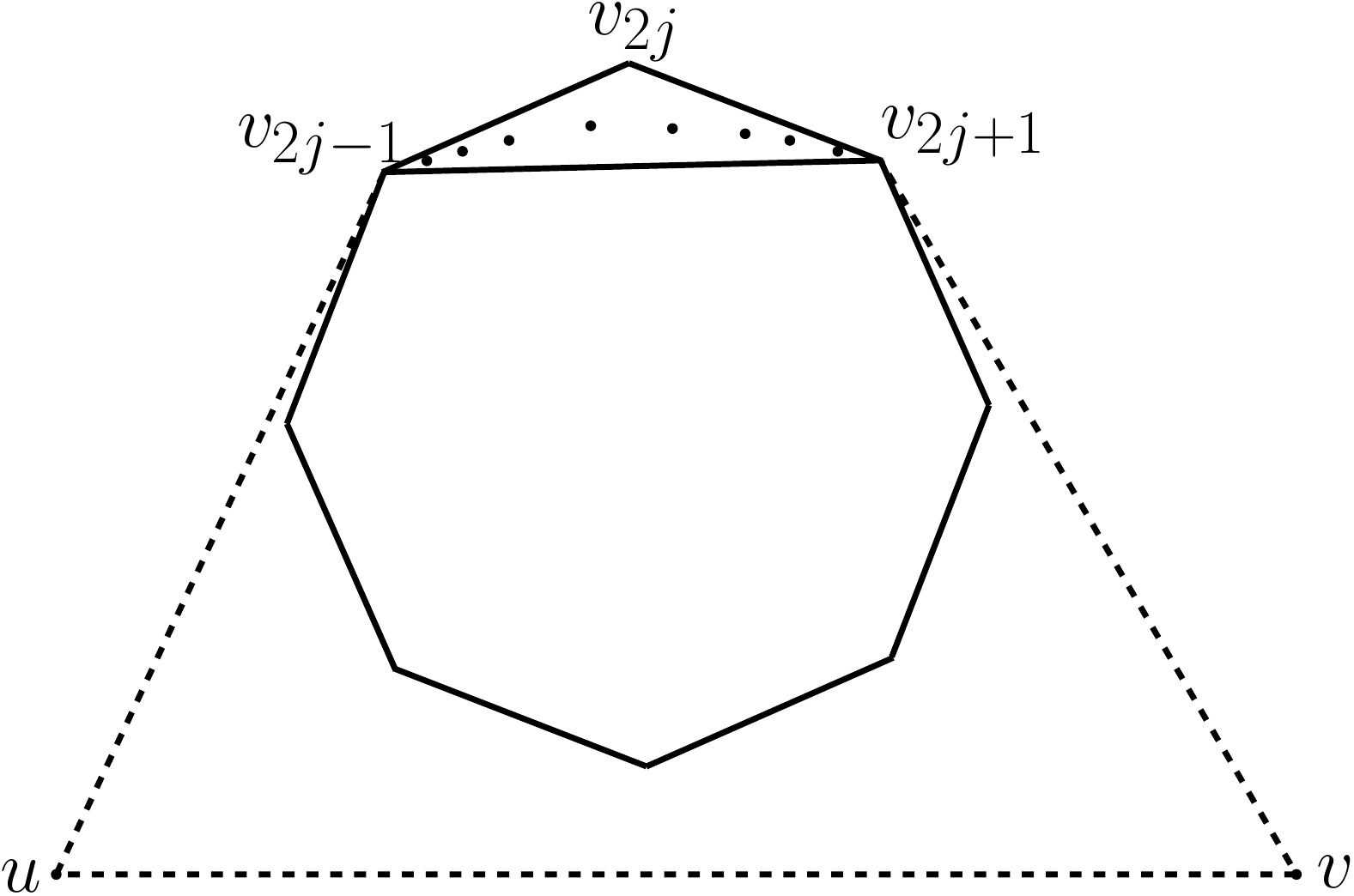}
  \caption{$u$, $v$ and the dotted points inside the triangle $\Delta
v_{2j-1}v_{2j}v_{2j+1}$ are the last $n+2$ points.}
  \label{fig:convhull}
\end{figure}
\remove{
\begin{wrapfigure}{r}{0.450\textwidth}
	
	\begin{center}
		\includegraphics[width=5cm]{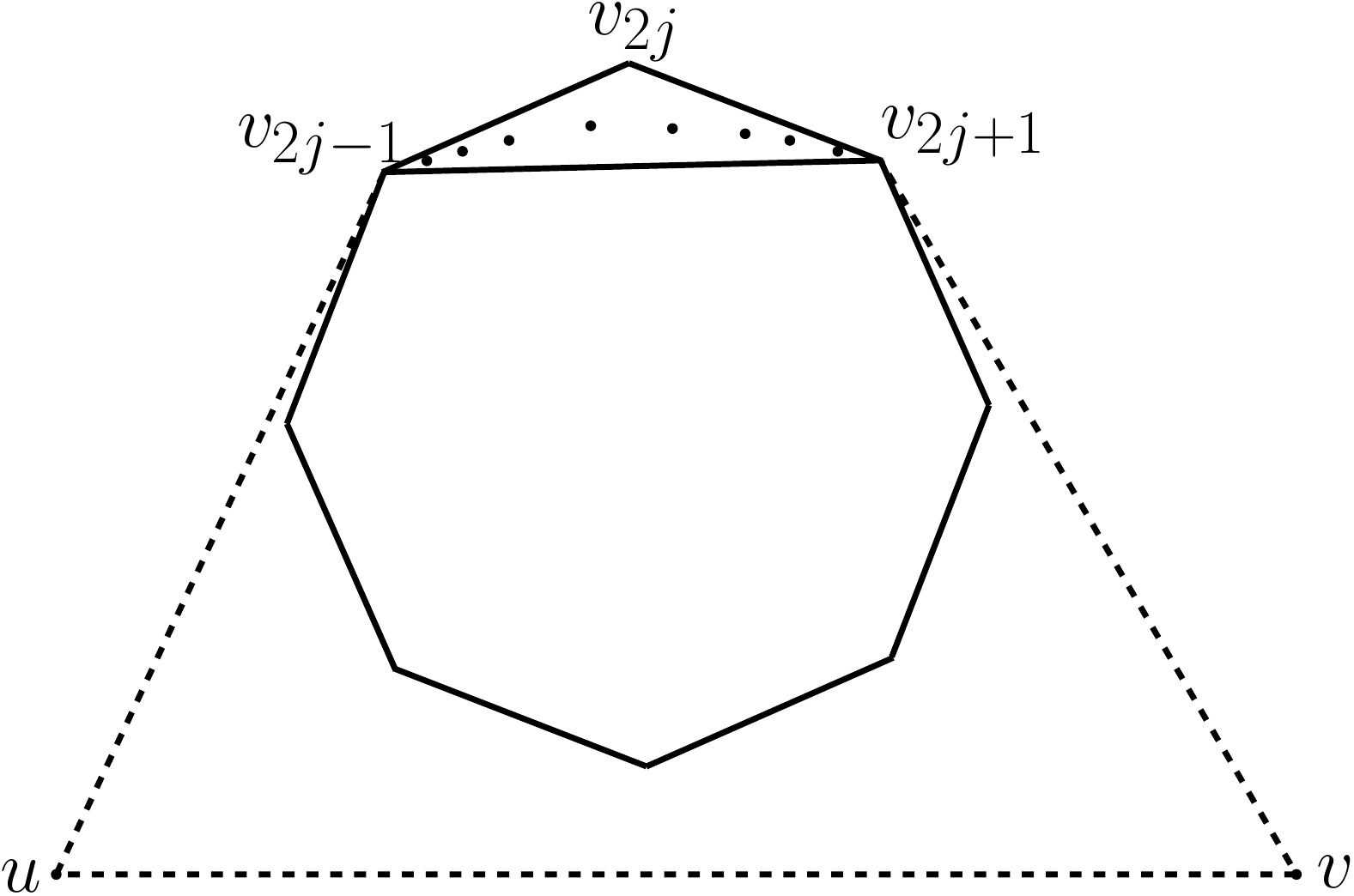}
		\caption{$u$, $v$ and the dotted points inside the triangle $\Delta
v_{2j-1}v_{2j}v_{2j+1}$ are the last $n+2$ points.}
		\label{fig:convhull}
	\end{center}

\end{wrapfigure}
}}
\begin{proof}
If there exists a randomized algorithm $\cA$ that distinguishes between $\size{\cCH(\cP)}=O(1)$ and $\size{\cCH(\cP)}=\Omega(n)$ with probability
$\rho$ and uses $o(n)$ bits, we can show the existence of a randomized protocol
that solves \ind and uses $o(n)$ bits.
Consider a regular convex polygon of $2n$ vertices, $V=\{v_i:1 \leq i \leq
2n\}$, as shown in Figure~\ref{fig:convhull}. We construct the input point set $\cP$
to $\cA$ by taking each $v_{2i-1}, 1 \leq 
i \leq n$. We also add $v_{2i}$ if and only if the 
$i$-th bit of Alice's input $x_i=1$. We send the current sketch to Bob.
Let $j$ be Bob's query index. Now another $n+2$ points are given as input to
$\cA$ such that $n$ of them are in convex position inside the triangle $\Delta
v_{2j-1}v_{2j}v_{2j+1}$ and the other two are such
that no vertices of $V \setminus \{v_{2j-1}, v_{2j}, v_{2j+1}\}$ can be on
$\cCH(\cP)$ as shown in Figure~\ref{fig:convhull}. Observe that the last $n+2$
points are put in such a way that if $v_{2j}$ is on $\cCH(\cP)$ then
$\size{\cCH(\cP)}=5$, otherwise,  $\size{\cCH(\cP)}=n+4=\Omega(n)$. Note that by
construction $v_{2j}$ is on $\cCH(\cP)$ if and only if $x_j=1$. Hence, $x_j=1$
if $\size{\cCH(\cP)}=5$ and $x_j=0$ if $\size{\cCH(\cP)}=\Omega(n)$.  
\end{proof}

\begin{figure}[!h]

\centering
%\begin{minipage}{0.5\textwidth}
\centering
\includegraphics[width=0.6\linewidth]{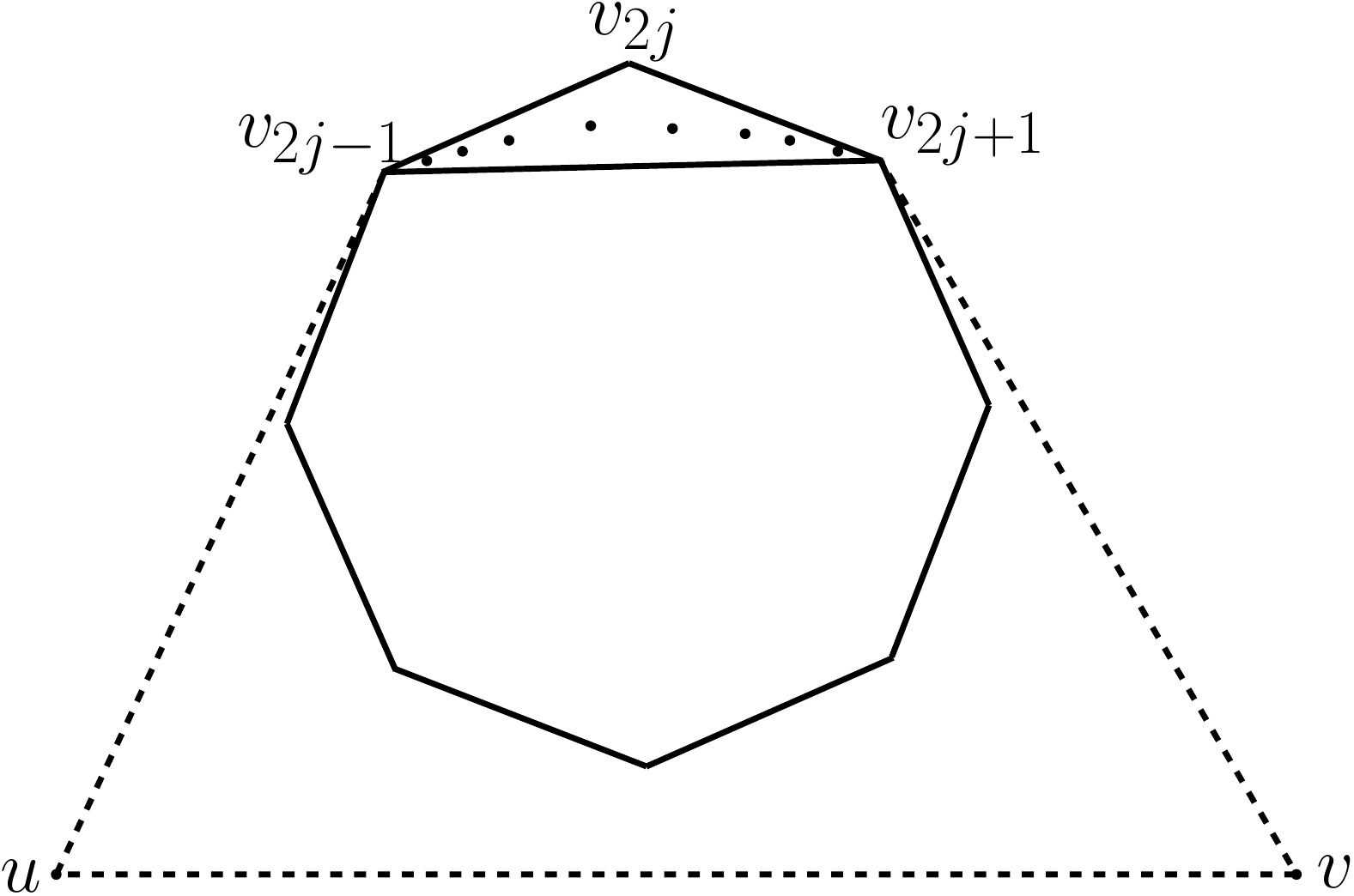}
  \caption{$u$, $v$ and the dotted points inside the triangle $\Delta
v_{2j-1}v_{2j}v_{2j+1}$ are the last $n+2$ points.}
  \label{fig:convhull}
\end{figure}

\subsubsection*{Lower bound for convex body approximation and geometric query}
\begin{figure}[!h]

%\end{minipage}%
%\begin{minipage}{0.5\textwidth}
\centering
\includegraphics[width=0.5\linewidth]{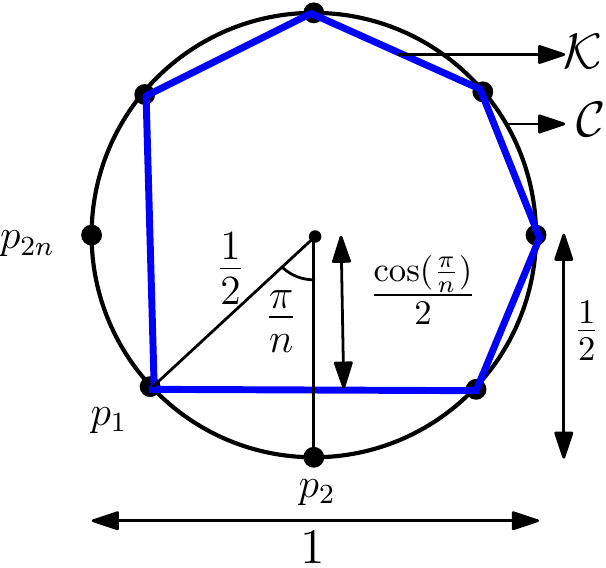}
\caption{Reduction idea for Theorem~\ref{theo:lb_convecapprox}.}
\label{fig:convappx}
%\end{minipage}%

\end{figure}
\begin{theo}
\label{theo:lb_convecapprox}
For every $\eps >0$, there exists a positive integer $n$ such that any (one-pass streaming) algorithm that $(\eps,\rho)$-approximates a convex polygon $\cK$, requires $\Omega\left(\frac{1}{\sqrt{\eps}} \right)$ bits of space, where $\cK$ is given as a stream of at most $2n$ straight lines in $\R^2$. 
\end{theo} 
\begin{proof}
Without loss of generality, assume that $0 < \eps <\frac{1}{2}$. Let $n=\ceil{ \frac{\pi}{2\sqrt{2\eps}}}$ and $\cA$ be a (streaming) algorithm, as stated, using
$o\left(\frac{1}{\sqrt{\eps}}\right)$ bits. Now we can design a protocol that solves \ind using space $o(n)$. See Figure~\ref{fig:convappx} for the following discussion.Alice and Bob know a circle $\cC$ of diameter 1 and $2n$ points $p_1,\ldots,p_{2n}$ placed evenly on its circumference. We process Alice's input ${\bf x}$ as follows. If $x_i=1$, then we give two line segments $\overline{ p_{2i-1}p_{2i}}$ and $\overline{p_{2i}p_{2i+1}}$ as inputs to $\cA$. If $x_i=0$, we give the line segment $\overline{p_{2i-1}p_{2i+1}}$ as an input to $\cA$. We send the current memory state of Alice to Bob.
\remove{
$\cK \subset \cC$, of diameter $1$ as shown in the Figure~\ref{fig:convappx}. $2n$ points $p_1,\ldots,p_{2n}$ are evenly placed on the circumference of $C$. The circle $C$ and $2n$ points are known to
both Alice and Bob. Now we process Alice's input ${\bf x}$ as follows. If $x_i=1$, then we give two line segments $\overline{ p_{2i-1}p_{2i}}$ and $\overline{p_{2i}p_{2i+1}}$ as inputs to $\cA$. If $x_i=0$, we give the line segment $\overline{p_{2i-1}p_{2i+1}}$ as an input to $\cA$. Now we send 
the current memory state of Alice to Bob.}
 Let $\cK$ be the actual convex polygon and $\cK'$ be the convex polygon generated by $\cA$ such that $d_H(\cK,\cK') \leq \eps$. Observe that $x_i=1$ implies $p_{2i} \in \cK$, i.e., there exists a point $q \in 
  \cK'$ such that $d(p_{2i},q)\leq \eps$, where $d(p,q)$ is the Euclidean distance between $p$ and $q$. If $x_i=0$, then $p_{2i}\notin \cK$ and $d(p_{2i}, \cK)= \min\limits_{q \in \cK}d(p,q)=\frac{1-
  \cos \left(\frac{\pi}{n}\right)}{2}=\sin^2 \left( \frac{\pi}{2n}\right) \geq \sin^2 \sqrt{2\eps}  > \eps$, as $0 < \eps < \frac{1}{2}$. 
  
Let $j$ be the query index of Bob. We report $x_j=1$ if and only if $\cC(p_{2j}) \cap \cK' \neq \phi$,
 where $\cC(p)$ denotes the circle of radius $\eps$ centred at $p$.
 \remove{
 \item[(b)] Let's consider the following promise version of \eql problem i.e P-\eql. Both inputs of Alice 
 and Bob i.e. ${\bf x}{\bf y} \in \{0,1\}^n$ and either ${\bf x}={\bf y}$ or there exists $set \frac{n}{4}$ consecutive indices $I \subset [n]$ such that $x_i \neq y_i,~\forall i \in I$. The objective is to decide 
 whether ${\bf x} ={\bf y}$. Formally $P-\eql({\bf x},{\bf y})=1$ if and only if ${\bf x}= {\bf y}$. Using 
 can be easily shown that the randomized private coin communication complexity of P-\eql is $\Omega(\log n)$. Details are in Appendix~\ref{}.
 
 Let $\cA$ be a (streaming) algorithm, as stated for $\eps=$, and uses $o(\log n)$ space. Now 
 design a protocol that solves P-\eql using $o(\log n)$ bits. 
 } 
\end{proof}

The proof for the lower bound of \geoqpoly is similar to the proof of Theorem~\ref{theo:lb_convecapprox}\remove{and is in Appendix~\ref{append:convbody}}.
\begin{figure}
  \centering
  \includegraphics[width=0.45\linewidth]{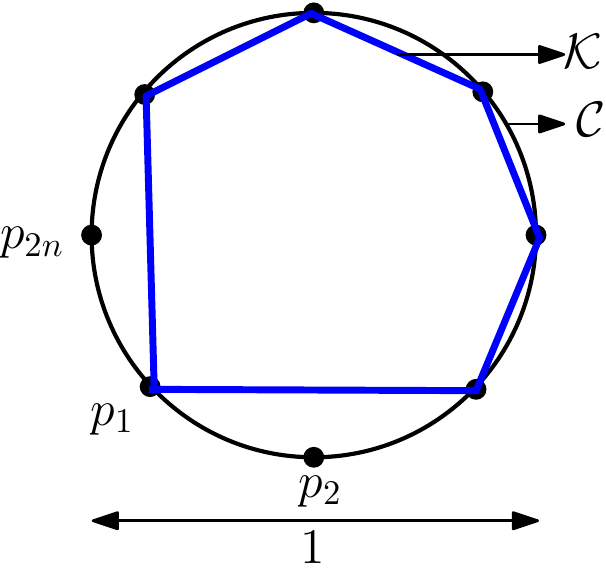}
\caption{Reduction idea for Theorem~\ref{theo:geoqconvpoly}.}
\label{fig:theo:geoqconvpoly}
\end{figure}

\begin{theo}
Every randomized one-pass streaming algorithm that solves \geoqpoly
with probability $1-\rho$,  uses $\Omega(n)$ bits of space.
\label{theo:geoqconvpoly}
\end{theo}
\begin{proof}
\remove{With out loss of generality, assume that $0 < \eps <\frac{1}{2}$. Let $n=\ceil{ \frac{\pi}{2\sqrt{2\eps}}}$ and }Let $\cA$ be a streaming algorithm, as stated in the Theorem, using
$o(n)$ bits. Now we can design a protocol that solves \ind using space $o(n)$ bits. See Figure~\ref{fig:theo:geoqconvpoly} for the following discussion. Alice and Bob know a circle $\cC$ of diameter 1 and $2n$ points $p_1,\ldots,p_{2n}$ placed evenly on its circumference. We process Alice's input ${\bf x}$ as follows. If $x_i=1$, then we give two line segments $\overline{ p_{2i-1}p_{2i}}$ and $\overline{p_{2i}p_{2i+1}}$ as inputs to $\cA$. If $x_i=0$, we give the line segment $\overline{p_{2i-1}p_{2i+1}}$ as an input to $\cA$. We send the current memory state of Alice to Bob.

 Let $\cK$ be the actual convex polygon. Observe that $x_i=1$ implies $p_{2i} \in \cK$. If $x_i=0$, then $p_{2i}\notin \cK$. Let $j$ be the query index of Bob. We report $x_j=1$ if and only if $\cA$ reports $p_{2j} \in \cK$.
\end{proof}

\subsection{Algorithms}
\label{sec:algo_conv}
\subsubsection*{Convex body approximation}
\noindent We state Dudley's result~\cite{Dudley74} and some observations about Hausdroff
distance that will be useful for the algorithm.

\begin{lem}\cite{Dudley74}\label{theo:dudley}
A convex body $\cK$ can be $\eps$-approximated by a
polytope $P$ with $1/{\eps^{(d-1)/2}}$ facets. $\eps$-approximation 
in this context means $d_H(\cK, P) \leq \eps$.
\end{lem}

\begin{obs}{\label{obs:subapx}}
Let $\cK_1,\cK_2,\cK_1'$ and $\cK_2'$ be convex bodies with $\cK_1 \subseteq \cK_1'$, 
$\cK_2 \subseteq \cK_2'$ and $d_H(\cK_1,\cK_1')\leq \eps_1, 
d_H(\cK_2,\cK_2')\leq \eps_2$. Then $d_H(\cK_1 \cap \cK_2, \cK_1' \cap \cK_2' )
\leq \max(\eps_1,\eps_2)$. 
\end{obs}

\begin{obs}{\label{obs:hosdfapx}}
Let $\cK,~\cK'$ and $~\cK''$ be convex bodies such that $d_H(\cK,\cK') \leq \eps_1$
and  $d_H(\cK', \cK'') \leq \eps_2$. Then, $d_H(\cK,\cK'')\leq \eps_1+\eps_2$. 
\end{obs}

Note that the approximated convex body in Lemma~\ref{theo:dudley} is always a
{\em superset} of the original one. Given a convex body $\cK$ in $\R^d$ and a
parameter $\eps > 0$, let $\cA$ be a non-streaming algorithm that outputs a
convex polytope $\cK'$ such that $d_H(\cK,\cK')\leq \eps$ and $\cA$ stores
$\frac{1}{\eps^{(d-1)/2}}$ facets.
Agarwal et al.~\cite{AgarwalHV04} used Bentley-Saxe's dynamization
technique~\cite{BentleyS80} to maintain the approximate extent measures of a
point set. We adapt these ideas for hyperplanes. Let a convex body $\cK$ be
given as a stream of hyperplanes in $\R^d$. $\cK$ is contained in a unit ball
$\cB$. Note that $\cK$ is the intersection of all hyperplanes in the stream.
The objective is to store a convex body $\cK'$ using sub-linear number of facets
that will $\eps$-approximate $\cK$, i.e., $d_H(\cK,\cK') \leq
\eps$.

We partition the processed stream of hyperplanes $\cH = <H_1,\ldots,H_n>$
into $t$  parts --- $\cH_1, \ldots, \cH_t$, $t \leq \ceil{\log n}$. For each $i \in [t]$, $\size{\cH_i} = 2^{r_i}$ for some non-negative integer $r_i < \ceil{\log n}$. We say $r_i$ is the rank of $\cH_i$. We maintain our data structure in a such a way that the rank of $\cH_i$ is equal to the rank of $\cH_j$ if and only if $i=j$. Let $C_i = \bigcap\limits_{H \in \cH_i}H$, be the convex polytope of hyperplanes 
in $\cH_i$. As we can not store $\cH_i$ or $C_i$, we maintain some approximation $C_i'$ of $C_i$ such that $d_H(C_i,C_i') \leq f(r_i)$. $f(r) = \eps/2r^2$ if $r \neq 0$ and $f(0)=0$. We store $C'_i$ and $r_i$. Let $\cC = \{C'_1,\ldots,C'_t\}$.

Now consider the situation when we have to process $H_{n+1}$. We set $\cH_{0} =
\{H_{n+1}\}$ and add $C'_{0}=C_{0}= H_{n+1} \cap \cB$ to $\cC$ as $f(0)$
approximation of $C_{0}$. If there exists $C'_i,C'_j \in \cC$ such that $r_i$ $=$ 
$r_j$ $=$ $r$ for some $r < \ceil{\log (n+1)}$, then using algorithm 
$\cA$, we find $C'$ as approximation of $C'_i \cap C'_j$ such that 
$d_H(C', C'_i\cap C'_j) \leq f(r+1)$. Note that the actual convex body corresponding to $ C'$ i.e., $\cH_i \cap \cH_j$ is of rank $r+1$. $\cC= (\cC \setminus \{C_i, C_j\})   \cup \{C'\}$. We repeat
the same process until all members of $\cC$ have distinct rank. Now our target
is that at the end, we should have each $C_i'$ as  approximation of $C_{i}$ such
that $d_H(C_{i},C'_i) \leq \eps$. As a result, we get $\cK'=
\bigcap\limits_{i=1}^{t}C'_{i}$ as an approximation of $\cK
=\bigcap\limits_{i=1}^{t}C_{i}$ such that $d_H(\cK,\cK') \leq \eps$ by
Observation \ref{obs:subapx}. Note that the number of times a $C_{i}$ is
approximated is at most its rank. Hence by Observation \ref{obs:hosdfapx}, we
have
%\begin{eqnarray*}
$$d_H(C_{i},C'_{i}) \leq \sum\limits_{l=1}^{r}f(l)  =
\sum\limits_{l=1}^{r}\frac{\eps}{2l^2} \leq \eps.$$ 
%\end{eqnarray*}

By Lemma~\ref{theo:dudley}, the amount of space used to store $C'_i$ is $O
\left (\frac{1}{f(r)^{(d-1)/2}} \right)$ where $r$ is the rank of $\cH_i$. We consider only for $r>0$, because there can be at most one $C_i$ of rank
$0$ and the corresponding approximation stores only $1$ facet. Hence, the total
amount of space we use is   
%\begin{eqnarray*}
$$1 + \sum_{l=1}^{t}  \size{C'_i} = O\left(\sum\limits_{l=1}^{\ceil{\log n}} \frac{1}{(f(l))^{(d-1)/2}}\right) =  O \left(\sum\limits_{l=1}^{\ceil{\log n}}\frac{1}{(\eps/2l)^{(d-1)/2}} \right) = O \left(\frac{\log ^{d} n}{\eps^{(d-1)/2}}\right)$$.
%\end{eqnarray*}

We summarize the above discussion in the following Theorem.
\begin{theo}
A convex body $\cK$, given as stream of $n$
hyperplanes, can be $\eps$-approximated deterministically by a polytope $P$
with $O\left(\frac{\log ^d n}{\eps^{(d-1)/2}}\right)$ facets.
\label{theo:convapx}
\end{theo}

\subsubsection*{Low dimensional \textsc{LP}} 

 \noindent Chan et al.~\cite{ChanC07} gave a multi-pass algorithm to find \emph{exact} solution to \lp, whose one-pass counterpart admits a lower bound of $\Omega(n)$ bits of space~\cite{GuhaM08}.
 
 In the streaming setting of \lp, constraints arrive as a stream of hyperplanes. Note that in \lp, the feasible region, i.e., intersection of all constraints (hyperplanes) is
a \emph{convex body}. Due to Theorem~\ref{theo:convapx}, given a set
of contraints as a stream, we can maintain $\eps$-approximation of the \emph{convex body}, i.e., the
 feasible region of the \lp using polylogarithmic~\footnote{$O(\frac{\log n}{\eps^{(d-1)/2}})$} space. The $\eps$-approximation along with the objective
function can be used to find an $\eps$-additive solution to \lp. 
Note that the $\eps$-approximtaion of the \emph{convex body} is a superset of original \emph{convex body}. So, the extreme point of the approximated convex body in the direction of the objective function 
vector may lie outside the feasible region. One can shift the facets of the
approximated convex body $\eps$ distance inwards to get a feasible solution
also. An added advantage is that the 
objective function need not be known beforehand. In summary, we have the
following Corollary to Theorem~\ref{theo:convapx}.\remove{ It is worth to look for $\eps$-additive solution to 
\lp as \emph{exact} solution to \lp admits a lower a bound of $\Omega(n)$ bits~\cite{ChanC07, GuhaM08}.}
\remove{
\begin{theo}\cite{GuhaM08}
Any one pass randomized streaming algorithm, that gives solves low dimensional {\em linear programming} with probability $2/3$, uses $\Omega(n)$ space. 
\end{theo}
}
\begin{coro}
There exists a one-pass deterministic streaming algorithm that outputs 
$\eps$-additive solution to \lp using $O\left(\frac{\log ^d n}{\eps^{(d-1)/2}}\right)$ space  for a fixed dimension $d$. The objective function may be revealed at the end of the stream of contraints.
\label{theo:lpapx}
\end{coro}

\subsubsection*{Property testing result for \textsc{GEOM-QUERY-CONV-POLY}}
\begin{figure}[!h]
  \centering
  \includegraphics[width=0.7\linewidth]{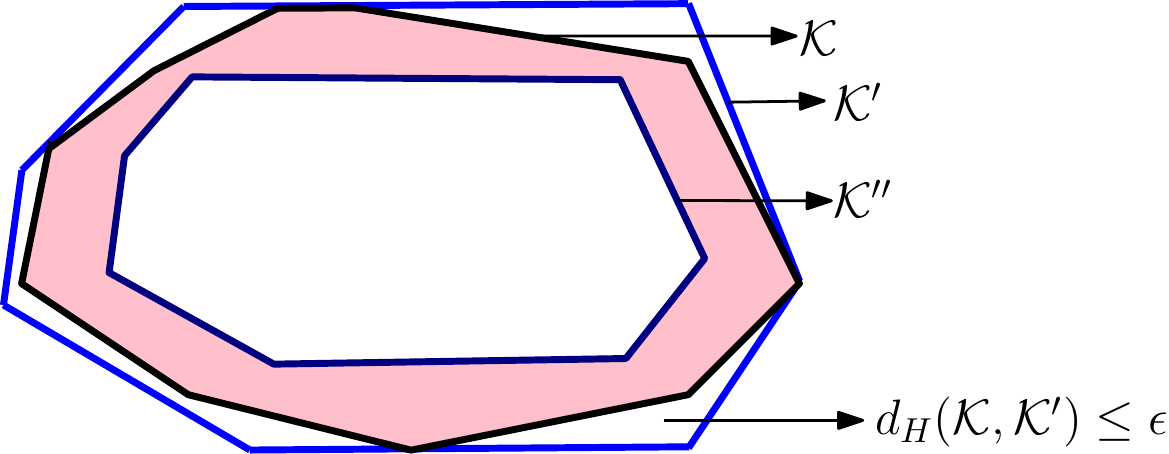}
  \caption{$\cK$ is the original convex body. $\cK'$ is the
$\eps$-approximation to $\cK$. $\cK''$ is shown as an $\eps$ shrink of $\cK$.
The region inside $\cK''$ and outside $\cK'$ are the zones for $q$ where we can
correctly answer.}
  \label{fig:geomqpoly}
  
\end{figure}

By the method discussed in Theorem~\ref{theo:convapx}, we
approximate $\cK$ by a convex body $\cK'$ such that $\cK \subseteq \cK'$ and
$d_H(\cK,\cK') \leq \eps $. By Theorem~\ref{theo:convapx}, the amount of space 
required to maintain $\cK'$ is $O\left(\frac{\log ^d n}{\eps^{(d-1)/2}}\right)$
facets. Note that at the end of the stream we have $\cK'$ as our sketch and 
 our objective is to answer correctly for query point $q$ if $d_H(\cK,\cK')\geq
\eps$. 
For \geoqpoly, we report (i) $q \in \cK$ if $q \in \cK'$ and $d_H(q,\cK') \geq
\eps$, (ii) $q \notin \cK$ if $q \notin \cK'$ and (iii) arbitrary answer,
otherwise. \remove{Let us assume that convex chain in \geoqchain is convex downward. We
report (i) $q$ is above $\cK$ if $q \in \cK'$ and $d_H(\cK,\cK')\geq \eps$, (ii)
$q$ is below $\cK$ if $q \notin \cK'$ and (iii) arbitrary answer, otherwise.}See Figure~\ref{fig:geomqpoly}. Summarizing the above discussion, we have the following Corollary to Theorem~\ref{theo:convapx}.
\begin{coro}
There exists a deterministic one-pass streaming algorithm that given a convex
polyhedron $\cK$ as a stream of hyperplanes  and a query point $q$ such that
$d_H(q, \cK)\geq \eps$, solves \geoqpoly  by storing  $O\left(\frac{\log ^d
n}{\eps^{(d-1)/2}}\right)$ facets.
\label{cor:geomqpoly}
\end{coro}

\section{Discrepancy problems}
\label{sec:discrepancy}

\remove{The only work prior to ours considering discrepancy in the streaming model has been the work by Agarwal et al.~\cite{AgarwalMPVZ06}. They defined \emph{discrepancy} in the context of spatial scan statistics and gave lower bounds and algorithmic results with respect to that. We stick to the conventional definition of both geometric and combinatorial discrepancy and our lower bound results are stronger than that of Agarwal et al.~\cite{AgarwalMPVZ06}. }

\noindent The proofs in this Section will require the notion of star discrepancy~\cite{bookkuipers}. In 
star discrepancy, all other problem specifications in the definition of discrepancy remain the same but 
each interval is constrained to have its left end point at $0$.

 Thus, the problem of \disc is to report $$\stgeo(\cP):=\sup\limits_{[0,q]\subseteq [0,1]}\size{q -\frac{n_{0q}}{n}},$$ and the problem of \stcdisc is to report  
 $$\stcol(\cP):=\stackrel[{I=[0,x]\subseteq [0,1)}]{\max}{} \size{R(I)-
B(I)},$$ which is same as $$\stcol(\cP)=\stackrel[{I_p=[0,p]: p \in
\cP}]{\max}{} \size{R(I_p)- B(I_p)}$$ because discrepancy values at points 
of the stream $\cP$ only matter. The following relations are known between the values of \emph{geometric discrepancy} $\geo(\cP)$ and \emph{color discrepancy} $\col(\cP)$ and their star variants. 
\begin{fact}~\cite{bookkuipers}
$\stgeo(\cP) \leq \geo(\cP) \leq 2\stgeo(\cP)$ and $\stcol(\cP) \leq \col(\cP) \leq 2 \stcol(\cP)$.
\label{fact:disc-relation}
\end{fact}
\begin{fact}~\cite{bookkuipers}
Let $x_1 < x_2 < x_3 < \ldots < x_n \in [0,1]$ be a sorted sequence of points in $\cP$. Then, the supremum in the definition of $\stgeo$ can be replaced by a maximum operation as $\stgeo(\cP) =
\max\limits_{i \in [n]}|x_i - \frac{2i-1}{2n}|$.
\label{fact:disc-star-alt-def}
\end{fact}
\subsection{Problem \textsc{GEOMETRIC-DISCREPANCY}}
%\subsection*{Lower bound} 

\begin{theo}\label{theo:lb_disc_geo}
For any $\eps>0$, there exists a positive integer $n$ such that any one-pass randomized streaming algorithm that outputs an $(\eps,\rho)$-additive solution to \gdisc for all streams of length $n$, requires $\Omega\left (\frac{1}{\eps} \right)$ bits of space.
\end{theo}

\begin{proof}
Let $\cP$ be the stream. We need the following claim that will be proved later.

\begin{cl}\label{lem:lb_disc}
For any $\eps>0$, there exists a positive integer $n$ such that any one-pass randomized streaming algorithm that outputs an approximate solution $D$ to \disc with probability $\rho$, such that $\stgeo(\cP)-\eps \leq D \leq 2\stgeo(\cP)+\eps$, requires $\Omega\left (\frac{1}{\eps} \right)$ bits of space, where $\cP$ is the input stream of length $2n$.
\end{cl}

Let there exist an algorithm as stated in Theorem~\ref{theo:lb_disc_geo} that returns $D'$ for $\eps$-additive error solution to \gdisc and uses space $o \left( \frac{1}{\eps}\right)$ bits. Observe that 
$\geo(\cP) -\eps \leq D' \leq \geo(\cP) -\eps$. Now using Fact~\ref{fact:disc-relation}, we can have the following. $D' \leq \geo(\cP)+\eps \leq 2\stgeo(\cP)+\eps$ and $D' \geq \geo(\cP)-\eps \geq \stgeo(\cP)-\eps$. So, we can report $D'$ as $D$, i.e., the solution to our \disc problem satisfying $\stgeo(\cP)-\eps \leq D \leq 2\stgeo(\cP)+\eps$. Note that we are using $o\left( \frac{1}{\eps}\right)$ bits, which contradicts Claim~\ref{lem:lb_disc}.  
\end{proof}

\remove{
\begin{rem}
\label{rem:multipass_disc}
Multipass lower bound: We can also do the reduction described in Lemma~\ref{lem:lb_disc} from \disj to any $p$-pass streaming algorithm, using space $s$, for computing $\geo(\cP)$. Observe that the 
induced protocol for \disj requires at most $2ps$ bits of communication. This implies
 $s =\Omega\left ( \frac{1}{\eps p}\right)$.\remove{ With similar arguement given in the proof of Theorem~\ref{theo:lb_disc_geo}, we can have the following. Any $p$-pass algorithm that returns $(\eps,\rho)$-multiplicative solution to \gdisc, requires $\Omega\left( \frac{1}{\eps p}\right)$ bits of space.}
\end{rem}
}
\begin{figure}[!h]
%\begin{minipage}{0.5\textwidth}
\centering
\includegraphics[width=0.8\linewidth]{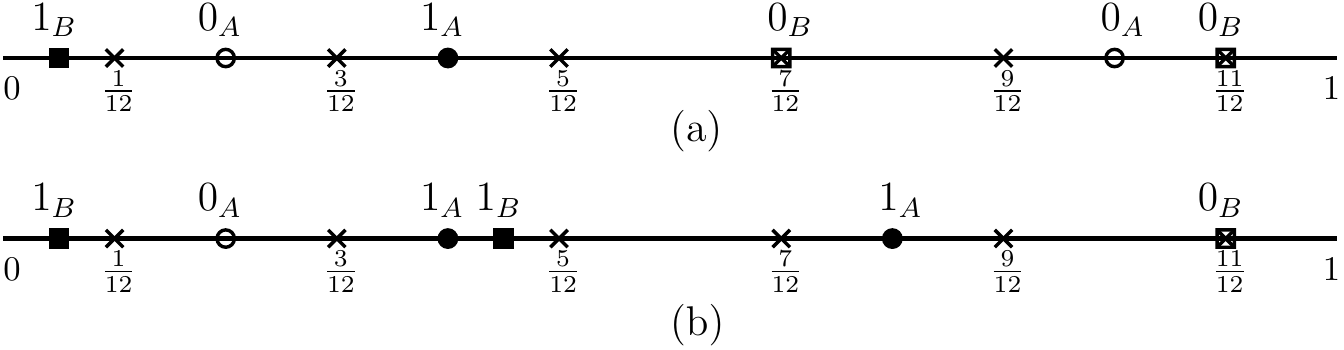}
  \caption{Reduction idea for Claim~\ref{lem:lb_disc}. Here $n=3$. In (a), Alice's input ${\bf x}=010$ and Bob's input ${\bf y}=100$; $\disj({\bf x},{\bf y})=1$; $\stgeo(\cP)=\frac{1}{12}$. In (b), ${\bf x}=011$ and ${\bf y}=110$; $\disj({\bf x},{\bf y})=0$; $\stgeo(\cP)=\frac{1}{6}+\frac{1}{24}$.}
  \label{fig:geodisc}
\end{figure}

\begin{proof}[Proof of Claim~\ref{lem:lb_disc}]

Let $n=\ceil{\frac{1}{32\eps}}$ and $\cA$ be a streaming algorithm that gives output, as stated, using $o\left( \frac{1}{\eps} \right)$ bits of space. We design a protocol 
for solving \disj using $o(n)$ bits. The idea is to generate points at varying intervals as per the inputs of Alice and Bob so that \disj can be solved by looking at the separation of the discrepancy values. 
We process Alice's bit vector ${\bf x}$ as follows. See Figure~\ref{fig:geodisc}, $1_A$ ($1_B$) denotes an input of $1$ for Alice (Bob) and $0_A$ ($0_B$) denotes an input of $1$ for Alice (Bob). 
For each $i \in [n]$, we give $z_i$ as input to $\cA$ such that
$z_i=\frac{4i-3}{4n}+\frac{1}{4n}$ if $x_i=0$ and $z_i=\frac{4i-3}{4n} -
\frac{1}{4n}$, otherwise. Let $Z_A = \cup_{i=1}^n z_i$. 
We send the current memory state to Bob. We process
Bob's input ${\bf y}$ as follows and give another $n$ inputs to $\cA$. We give input 
$z_{n+i}=\frac{4i-1}{4n}$ if $y_i=0$ and
$z_{n+i}=\frac{4j-3}{4n} - \frac{1}{8n}$ if $y_i=1$. Let $Z_B = \cup_{i=1}^n z_{n+i}$ and 
$Z=Z_A \cup Z_B$. In total, we have given $Z$ having  
$2n$ inputs to $\cA$. Let $Z_s = <z_1',\ldots,z_{2n}'>$, be the sorted sequence, in increasing fashion, of the points in $Z$. Recalling Fact~\ref{fact:disc-star-alt-def}, note that $\stgeo(\cP)=\max\limits_{i \in [2n]}
D_i$, where $ D_i=  \size{\frac{2i-1}{4n}-z_i'}$. Let $J=\{j:y_j=1\}$ be the set of indices where Bob has an input of 1. Let  $A(j), B(j)$, $j \in J$, be the indices of the input in $Z_s$ corresponding to $z_j$ (Alice) and
$z_{n+j}$ (Bob), respectively. Note that $B(j)=A(j)+1$ if $x_j=1$ and $B(j)=A(j)-1$, otherwise . One can
see that $D_i \leq \frac{1}{4n}$, $\forall i \in[2n]~\mbox{and}~ i \neq A(j), B(j); j \in J$. If $x_j = 1$, $B(j)=A(j) + 1$,
$D_{A(j)}  = \frac{1}{4n}$  and $D_{B(j)} =\frac{1}{2n}+\frac{1}{8n}$. If $x_j = 0$, $B(j)=A(j) - 1$, $D_{A(j)}  = \frac{1}{4n}$ and $D_{B(j)} = \frac{1}{8n}$. Hence, if $\disj({\bf x,y})=0$, there exists an index $i\in [n]$ such that $x_i=y_i$, then $\stgeo(\cP) = \frac{1}{2n}+\frac{1}{8n}$ and in this case $\cA$ reports at least $\frac{1}{2n}+\frac{1}{8n}-\eps$, i.e., $19 \eps$. If $\disj({\bf x,y})=1$, then $\stgeo(\cP) \leq \frac{1}{4n}$ and in this case $\cA$ reports at most $2\stgeo(\cP)+\eps=\frac{1}{2n}+\eps$,  i.e.,  $17\eps$.  So, 
we report $\disj({\bf x,y})=1$ if and only if $\cA$ reports at most $17\eps$.
\end{proof}

We have been able to design deterministic one (multi) pass streaming algorithm for
$\eps$-additive (multiplicative) solution to \gdisc using bucketing technique. The result is stated in
the following Theorem\remove{, whose proof is in Appendix~\ref{append:disc}}. 
\remove{ as points
in one dimension follow a total order. In $\R^2$, 
consider a point set $\cP$ with no two points having the same
$x$ or $y$ coordinate. In this case \disc
 is denoted and defined as $D_2^*(\cP)= \sup\limits_{0 < x, y <
1}\size{xy-\frac{n_{xy}}{n}}$
where $n_{xy}=\size{ \{(x_i,y_i) \in [0, x] \times [0, y], i \in [n]\}}$.
\begin{obs}
%Two dimensional $\disc$ can be defined as 
$D_2^*(\cP) = \max \limits_{p_i\in \cP} \max \limits_{p_j \in \cP} \max \left
(\size{x_iy_j - \frac{n_{x_iy_j}}{n}}, \size{x_iy_j - \frac{n_{x_iy_j} - 2}
{n}} \right) $.
\end{obs}
For \disc in $\R^2$, the objective is to determine $D_2^*(\cP)$ where $n$ is
unknown until end the of the stream. Following is the result for a special case
where minimum distance between any pair of points is some $\delta > 0$ and $n
\geq \frac{c}{\delta^2}$ for some constant $c$. This essentially
means that the algorithm works for ``uniform'' point sets.  
For points in two dimensions, 
the bucketing technique does not generalize. We need a separation condition 
on the points to make a careful analysis of a bucketing technique using 
packing arguments. For want of space, the proofs are in Appendix~\ref{append:disc}.  
} 
\begin{theo}
\label{theo:disc_ub}
There exists a one-pass deterministic streaming algorithm for
$\eps$-additive solution to \gdisc using $O(\frac{1}{\epsilon})$
space, where $\epsilon\in(0,1)$ is an input parameter.
\remove{
 \item[(b)] There exists an one-pass deterministic streaming algorithm that outputs
$\eps$-additive solution to \disc in $\R^2$ for a constrained point set $\cP$ and
uses $O(\frac{1}{\epsilon^2})$ space; $\epsilon\in(0,1)$ is an input parameter
such that $\epsilon >
\frac{6}{\sqrt{n}}$.} 
\label{theo:discrepancy-one-two-algo}
\end{theo}
\begin{proof}\remove{[Proof of Theorem~\ref{theo:discrepancy-one-two-algo}]}

We use bucketing technique to solve \gdisc in $\R$. Given an $\epsilon\in
(0,1)$, we create $ \lceil \frac{1}{\epsilon}
\rceil$ buckets. $\cB= \{B_i=[2(i-1)\eps, \min(2i\eps,1)):
i\in[\ceil{\frac{1}{\eps}}]\}$ be
the set of buckets. We also maintain $\ceil {\frac{1}{\epsilon}}$ counters,
i.e., $c_i, i\in[\ceil{\frac{1}{\eps}}]$ such that $c_i$ maintains the number of
points in $B_i$. On receiving a point in the
stream, we only increase the count of the corresponding counter of the bucket.

  At the end of the stream, we know the value of $n$ and the values of $c_i$'s, $i \in
\ceil{\frac{1}{2\eps}}$. Let $C_i$ denote the number of points in
$\stackrel[j=1]{i}{\cup} B_j$ i.e., $C_i =
\sum\limits_{j=1}^{i}c_j$. Note that only $c_i$'s (and $C_i$'s) are maintained,
not exact coordinate of points. We take $y_i = (2i-1)\frac{\eps}{2}$ as the representative coordinate for
each point in $B_i$. So, the amount of space we are using is 
$O\left( \frac{1}{\eps} \right)$.

  For buckets $B_i$ and $B_j$, $i \leq j$, consider any interval $[a,b]\subseteq [0,1]$ that spans from $B_i$ to $B_j$, i.e., $a\in B_i$ and $b \in B_j$. So, if we know the exact value of the number of points present in $[a,b]$, i.e., $n_{ab},$ We can report $\size{\size{[y_i,y_j]}-\frac{n_{ab}}{n}}$ as an $\eps$-additive solution to $\size{\size{[a,b]}-\frac{n_{ab}}{n}}$ as $\size{y_i-a}, \size{y_j-b}\leq \frac{\eps}{2}$. But we do not know exact $n_{ab}$. However, $C_{i-1}+1 \leq n_{ab} \leq C_j$, 
where $C_0=0$, as $[a,b]$ spans from $B_i$ to $B_j$. So, 
$$
  	\max\limits_{C_{i-1}+1 \leq n_{ab} \leq C_j}\left(\size{\size{[y_i,y_j]}-\frac{n_{ab}}{n}}\right)
  	=\max\left( \size{\size{[y_i,y_j]}-\frac{C_{i-1}+1}{n}}, \size{\size{[y_i,y_j]}-\frac{C_{j}}{n}}\right)
$$ 
  is an $\eps$-additive approximate solution to
   $\max\limits_{a\in B_i, b\in B_j}\size{\size{[a,b]}-\frac{n_{ab}}{n}}$.
   
    Observe that $\geo(\cP)=\max\limits_{1\leq i \leq j \leq \ceil{\frac{1}{\eps}}}\max\limits_{a\in B_i, b\in B_j}\size{\size{[a,b]}-\frac{n_{ab}}{n}}$. 
    Hence, we report 
    $$
    	\max\limits_{1\leq i \leq j \leq \ceil{\frac{1}{\eps}}}\max\left( \size{\size{[y_i,y_j]}-\frac{C_{i-1}+1}{n}}, \size{\size{[y_i,y_j]}-\frac{C_{j}}{n}}\right)
    $$ 
    as our $\eps$-additive solution to $\geo(\cP)$. 
  \remove{
  We want to report $D_r = \size{x_r - \frac{2r-1}{2n}}$, where
$x_r$ is the data whose rank is $r$ in $\cP$. Find $i$ such that $C_{i-1} < r
\leq {C}_{i}$. We report $D_r'=\size{y_i -\frac{2r-1}{n}}$. Now as $|x_r -y _i|
\leq \epsilon$, 
observe that  $| D_r - {D_r}^{*}| \leq  \epsilon$. Trivially, by putting
$r=1(1)n$, we can find $D_{r}$ and report the maximum among them as $D$
such that $\size{D_1(\cP) - D_1^*(\cP) } \leq \eps$. It will take $O(n)$
time after the end of the stream. To solve it efficiently, the key observation
is that the function $|x-
\frac{i}{n}|, i \in[a, b]$, attains its maximum at either  $x=a$ or $x=b$ for
fixed $x$ and $n$. Hence for each bucket(say $i^{th}$) we compute $D_{r}$ only
for $r= C_{i-1} + 1$ and $r=C_i$ and report maximum among them
as output, which will take $O(1/\epsilon)$ time as opposed to trivial $O(n)$
time after the end of the stream. The amount of space we are using here is 
$O\left( \frac{1}{\eps} \right)$.}
This concludes the proof.  
\end{proof}
\remove{\begin{rem}
A similar bucketing technique as above, can be applied to find $\eps$-additive ($\eps$-multiplicative) solution to \gdisc in two dimension, under some condition, by using space $O\left(\frac{1}{\eps^2} \right)$. The details are in Appendix.
\end{rem}}
\subsection{Problem \textsc{COLOR-DISCREPANCY}}
\label{ssec:colordisc}
\remove{
 \begin{obs}\label{obs:cdisc1}
 $D_1^{*c}(\cP)=\stackrel[{I_p=[0,p): p \in \cP}]{\max}{} \size{R(I_p)- B(I_p)}$.
 \end{obs}
 \begin{obs}\label{obs:cdisc2}
 $D_1^{*c}(\cP) \leq D_1^c(\cP) \leq 2D_1^{*c}(\cP)$.
 \end{obs}
}
\remove{\begin{lem}\label{lem:lb_cdisc}
 Any one-pass streaming algorithm that returns $D$ with probability such that $ (1-\eps)D_1^{*c}(\cP)\leq D \leq 2(1+\eps)D_1^{*c}(\cP) $, uses $\Omega(n)$ bits of space, where $\cP$ is the input stream of length $n$ and $0 < \eps <\frac{1}{5}$.
\end{lem}}

 \begin{theo}\label{theo:lb_cdisc}
Any one-pass streaming algorithm that returns $(\eps,\rho)$-multiplicative solution to \cdisc, uses $\Omega(n)$ bits of space, where $\cP$ is the input stream and $0 < \eps <\frac{1}{5}$. 
\end{theo}
\begin{figure}[!h]
\centering
\includegraphics[width=0.8\linewidth]{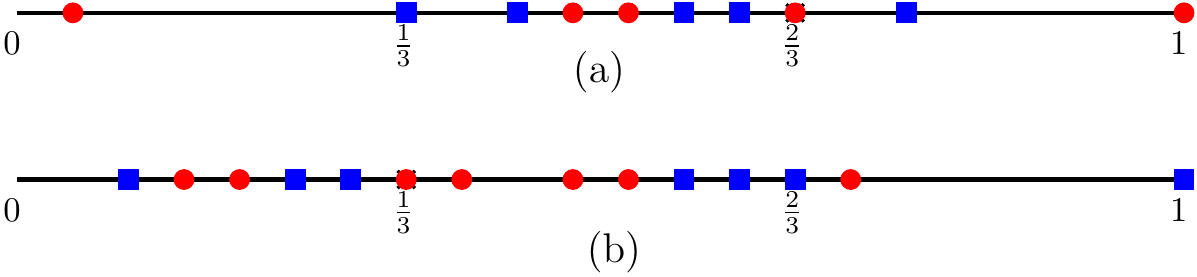}
\caption{Reduction idea for Claim~\ref{lem:lb_cdisc}. Here $n=3$. In (a), Alice's input ${\bf x}=100$ and Bob's input ${\bf y}=010$; $\disj({\bf x},{\bf y})=1$; $\stcol(\cP)=2$. In (b), ${\bf x}=011$ and ${\bf y}=110$; $\disj({\bf x},{\bf y})=0$; $\stcol(\cP)=3$.}
\label{fig:combdisc}
\end{figure}
\begin{proof}
We need the following claim where $\cP$ be the stream.  
\begin{cl}\label{lem:lb_cdisc}
 Any one-pass streaming algorithm that outputs an approximate solution $D$ to \stcdisc with probability $\rho$ such that $ (1-\eps)\stcol(\cP)\leq D \leq 2(1+\eps)\stcol(\cP) $, uses $\Omega(n)$ bits of space, where $\cP$ is the input stream of length $n$ and $0 < \eps <\frac{1}{5}$.
\end{cl}
Let there exist an algorithm as stated in Theorem~\ref{theo:lb_cdisc} that outputs $D'$ for \cdisc and uses space of $o(n)$ bits. Now using Fact~\ref{fact:disc-relation}, we have $D' \leq (1+\eps)\col(\cP) \leq 2(1+\eps)\stcol(\cP)$ and $D' \geq (1-\eps)\col(\cP) \geq (1-\eps) \stcol(\cP)$. So, we can report $D'$ as our $D$, i.e., the approximate solution to our \stcdisc satisfying $(1-\eps)\stcol(\cP)\leq D \leq 2(1+\eps)\stcol(\cP) $. Note that we are using $o(n)$ bits, which contradicts Claim~\ref{lem:lb_cdisc}. 
\end{proof}
\begin{rem}
\label{rem:multipass_disc_color}
Multipass lower bound: Using a similar line of arguemnet of Remark~\ref{rem:klee_lb}, we can have the followings.
\begin{itemize}
\item Any $p$-pass algorithm that computes $\eps$-multiplicative solution to \cdisc (\gdisc), requires $\Omega\left( \frac{n}{p} \right)$ bits of space, where $0< \eps <\frac{1}{5}$.
\item Any $p$-pass algorithm that computes $\eps$-additive solution to \cdisc (\gdisc), requires $\Omega\left( \frac{1}{\eps p} \right)$) bits of space, where $0< \eps < 1$.
\end{itemize} 
\end{rem}
\begin{proof}[Proof of Claim~\ref{lem:lb_cdisc}]
We show a reduction from \disj. Let $\cA$ be an algorithm that solves
correctly \stcdisc, as stated, with probability $2/3$ and uses $o(n)$ bits of space. Now
we can design a protocol by suitably placing ``red'' and ``blue'' points and looking for separation of discrepancy values to solve \disj. We process each bit of Alice's input
${\bf x}$ as follows. See Figure~\ref{fig:combdisc}. If $x_i =1$, we give inputs $\frac{i-1}{n} + \frac{1}{7n}$
and $ \frac{i}{n}$ labeled as ``red'' and ``blue'', respectively to $\cA$.
Otherwise, we give points $\frac{i-1}{n} + \frac{2}{7n}$ and $ \frac{i}{n}$
labeled as ``blue'' and ``red'', respectively as inputs. We send the current
memory status of $\cA$ to Bob. Bob processes his input ${\bf y}$ as follows. If $y_i=1$, Bob gives four
inputs to $\cA$: $\frac{i-1}{n} + \frac{3}{7n}$ and $\frac{i-1}{n} + \frac{4}{7n}$  both labeled as ``red''; $\frac{i-1}{n} + \frac{5}{7n}$ and $\frac{i-1}{n} + \frac{6}{7n}$ both labeled as ``blue''. If $y_i=0$, Bob does nothing. As discussed at the begining of this Section, the discrepancy values at the input 
points only matter. By construction of the input instance of \stcdisc, each point in $\cP$ is in
one of the forms: $\frac{i-1}{n} + \frac{j}{5n}$ for some $i \in [n], 0 \leq j \leq 6$. Recall that $I_p=[0,p]$. Observe that, $\size{R(I_p)-B(I_p)} = 0$ if
$p = \frac{i}{n}$ for some $i$; $\size{R(I_p)-B(I_p)} = 1$ if $p= \frac{i-1}{n}+
\frac{1}{7n}$ or $p= \frac{i-1}{n}+ \frac{2}{7n}$ for some $i$. Let $J=\{k:y_k = 1\}$. Only for $i \in J$,
we have $I_p$ such that $p= \frac{i-1}{n}+ \frac{j}{7n}$, where $3 \leq j \leq 6$. 

Observe that if $x_j = 1$,  then
\begin{eqnarray*}
\size{R(I_p)-B(I_p)} &=& 2~\mbox{for}~ p=\frac{j-1}{n}+\frac{3}{7n},\\
 \size{R(I_p)-B(I_p)} &=& 3~\mbox{for}~ p=\frac{j-1}{n}+\frac{4}{7n}, \\
 \size{R(I_p)-B(I_p)} &=& 2 ~\mbox{for}~ p=\frac{j-1}{n}+\frac{5}{7n},\\
 ~\mbox{ and}  \size{R(I_p)-B(I_p)} &=& 1 ~\mbox{for}~ p=\frac{j-1}{n}+\frac{6}{7n}.
\end{eqnarray*}
 If $x_j = 0$, then
 \begin{eqnarray*}
 \size{R(I_p)-B(I_p)} &=& 0 ~\mbox{for}~  p=\frac{j-1}{n}+\frac{3}{7n},\\
 \size{R(I_p)-B(I_p)} &=& 1 ~\mbox{for}~  p=\frac{j-1}{n}+\frac{4}{7n}, \\
 \size{R(I_p)-B(I_p)} &=& 0 ~\mbox{for}~  p=\frac{j-1}{n}+\frac{5}{7n},\\ 
  ~\mbox{ and} \size{R(I_p)-B(I_p)} &=& 1 ~\mbox{for}~  p=\frac{j-1}{n}+\frac{6}{7n}.
 \end{eqnarray*}
   
 If $\disj ({\bf x}, {\bf y})=0$, there exists an index $i$ such that $x_i=y_i=1$, then 
 $\stcol(\cP)=3$  and in this case $\cA$ returns at least $(1-\eps)\stcol(\cP)$, i.e., more than $\frac{12}{5}$.  If $\disj ({\bf x}, {\bf y})=1$, then 
 $\stcol(\cP)=1$  and in this case $\cA$ returns at most $2(1+\eps)\stcol(\cP)$, i.e., less than $\frac{12}{5}$. Hence, we report $\disj({\bf x},{\bf y})=1$ if and only if $\cA$ gives output less than $\frac{12}{5}$.
 \end{proof}
\remove{
 From the proof of Theorem~\ref{lem:lb_cdisc}, one can observe that $x_j=1$ if
$D_1^{*c}(\cP)=2$ and $x_j=0$ if $D_1^{*c}(\cP)=1$. There is a gap between 1
and 2. So, if we can give a $1 \pm \eps$-approximation factor algorithm,
$\epsilon \le \frac{1}{3}$, for \cdisc with probability $2/3$, then we can
solve \ind with probability $2/3$. Hence, we have the following lemma.
 \begin{lem}
 Any one-pass randomized streaming algorithm that returns
$D_1^{*c}(\cP)$ within a  $1 \pm \eps$ approximation factor, $\eps <
\frac{1}{3}$, with probability $2/3$, uses $\Omega(n)$ bits of space.
 \end{lem}
 } 
 \subsection{Problem \textsc{COLOR-DISCREPANCY} for sorted sequence}
 \label{ssec:cdisc_sort}

 By Theorem~\ref{theo:lb_cdisc}, approximating $\cdisc$ needs $\Omega(n)$
 bits. But if the stream $\cP$ arrives in a sorted order, we can compute $\col(\cP)$ using $O(1)$ space.  
 \remove{
 \subsection*{How to compute $D_1^{*c}(\cP)$?} 
 Maintain number of red and blue points  seen so far i.e $R$ and $B$ respectively.
 Also maintain another two variables i.e. $max$ and $min$ to maintain $D_1^{*c}$ of
  the points seen so far. On receiving an input increment $R$ or $B$ accordingly and 
  then update $max$ and $min$ by
  $\max (max,R-B)$ and $\min(min,R-B)$ respectively. At the end we report $\max(max,min)$, which is the
  required $D_1^{*c}(\cP)$ of the sorted stream.
  \subsection*{How to compute $D_1^{c}(\cP)$?} 
For a constant space algorithm to find $D_1^c(\cP)$ we need Observation~\ref{obs:cdisc_sort} followed 
by Lemma~\ref{lem:cdisc_sort}.
}
\begin{theo}
\label{theo:cdisc_sort}
\cdisc can be solved exactly by a one-pass deterministic streaming algorithm using $O(1)$ space when 
$\cP$ is sorted.
\end{theo}
\begin{proof}
Let $I_{opt}=[p,q]$ be an interval where $\col(\cP)=\size{R(I_{opt})-B(I_{opt})}$ is optimized.
Then $p,q \in \cP$ and $D_c(\cP)\neq 0$, i.e., $R(I_{opt}) \neq B(I_{opt})$. Also if 
$R(I_{opt})>B(I_{opt})$ then $p,q$ are red points and if $R(I_{opt})<B(I_{opt})$ then
$p,q$ are blue points. To establish the theorem, we need the following Claim\remove{, which we 
prove in Appendix~\ref{append:disc}}.
\begin{cl}
\label{lem:cdisc_sort}
$\col(\cP)=\max\limits_{p\in \cP}\left(R\left([0,p]\right)-B\left([0,p]\right)\right)- \min\limits_{q \in \cP}\left(R\left([0,q]\right)-B\left([0,q]\right)\right) +1$.
\end{cl}
The algorithm keeps track of the number of red (denoted as $\#R$) and blue (denoted as $\#B$) points seen so far. It also maintains aother two variables --- $max$ and $min$. On receiving an input, increment $\#R$ or $\#B$ accordingly and then update $max$ with $\max (max,\#R-\#B)$ and $min$ with $\min(min,\#R-\#B)$. Observe that $\col(\cP)=max-min+1$ by Claim~\ref{lem:cdisc_sort}.
\end{proof}
\small{
\begin{proof}[Proof of Claim~\ref{lem:cdisc_sort}]
\begin{eqnarray*}
\col(\cP) &=& \max\limits_{I\subset[0,1]}\size{R(I)-B(I)}\\
&=& \max\limits_{p,q \in\cP}\size{R\left([p,q]\right)-B\left([p,q]\right)}\\
&=& \max\Big(
 \max\limits_{p,q \in \cP, R\left([p,q]\right)>B\left([p,q]\right)}\left( R\left([p,q]\right)-B\left([p,q]\right)\right), \\
&&\quad\quad\quad\quad\quad\quad \max\limits_{p,q \in \cP, B\left([p,q]\right)>R\left([p,q]\right)} \left( B\left([p,q]\right)-R\left([p,q]\right)\right)\;\;\;\;
 \Big)\\
 &=& \max\Big(
 \max\limits_{p,q \in \cP, R\left([p,q]\right)>B\left([p,q]\right)}\left( \left( R\left([0,q]\right)-R\left([0,p]\right)+1\right)-\left(B\left([0,q]\right)-B\left([0,p]\right)\right)\right),  \\
 && \quad\quad\quad\quad\quad\quad
 \max\limits_{p,q \in \cP, B\left([p,q]\right)>R\left([p,q]\right)}\left( \left( B\left([0,q]\right)-B\left([0,p]\right)+1\right)-\left(R\left([0,q]\right)-R\left([0,p]\right)\right)\right)
 \Big)\\
 \remove{
 &=& \max\limits_{p,q \in \cP}\size{(R([0,q])-R([0,p]))-(B([0,q])-B([0,p]))} + 1\\
  &=& \max\limits_{p,q \in \cP}\size{(R([0,q])-B([0,q]))-(R([0,p])-B([0,p]))} + 1\\
  &=& \max_{p \in \cP}(R([0,p])-B([0,p]))-\min_{p \in \cP}(R([0,p])-B([0,p])) + 1}\\
%\remove{\col(\cP) &=& \max\limits_{I\subset[0,1]}\size{R(I)-B(I))}\\
%&=& \max\limits_{p,q \in\cP}\size{R\left([p,q]\right)-B\left([p,q]\right)}\\
%&=& \max\left(
%\max\limits_{p,q \in \cP, R\left([p,q]\right)>B\left([p,q]\right)}\left( R\left([p,q]\right)-B\left([p,q]\right)\right), \max\limits_{p,q \in \cP, B\left([p,q]\right)>R\left([p,q]\right)} \left( B\left([p,q]\right)-R%\left([p,q]\right)\right)
% \right)\\
% &=&\max\left(
% \max\limits_{p,q \in \cP, R\left([p,q]\right)>B\left([p,q]\right)}\left( \left( R\left([0,q]\right)-R\left([0,p]\right)+1\right)-\left(B\left([0,q]\right)-B\left([0,p]\right)\right)\right),  
% \max\limits_{p,q \in \cP, B\left([p,q]\right)>R\left([p,q]\right)}\left( \left( B\left([0,q]\right)-B\left([0,p]\right)+1\right)-\left(R\left([0,q]\right)-R\left([0,p]\right)\right)\right)
% \right)\\}
 &=& \max\limits_{p,q \in \cP}\size{(R([0,q])-R([0,p]))-(B([0,q])-B([0,p]))} + 1\\
  &=& \max\limits_{p,q \in \cP}\size{(R([0,q])-B([0,q]))-(R([0,p])-B([0,p]))} + 1\\
  &=& \max_{p \in \cP}(R([0,p])-B([0,p]))-\min_{q \in \cP}(R([0,q])-B([0,q])) + 1
\end{eqnarray*}
\end{proof}
}
\bibliographystyle{alpha}
\bibliography{Lowerbound}

\newpage
\appendix

\end{document}